\documentclass[showpacs,superscriptaddress,amsfonts,amsmath,floatfix,aps,10pt,pra,twocolumn]{revtex4-1}
\usepackage[T1]{fontenc}
\usepackage[latin2]{inputenc}
\usepackage{graphicx}
\usepackage{epsfig}

\input{Qcircuit}

\newcommand{\re}{\mathrm{Re}\, }
\newcommand{\im}{\mathrm{Im}\,}
\newcommand{\I}{{\rm{i}}}
\newcommand{\D}{{\rm{d}}}

\newcommand{\tr}{\operatorname{tr}}

\newcommand{\dss}{\displaystyle}
\newcommand{\nn}{\nonumber}

\newcommand{\be}{\begin{equation}}
\newcommand{\ee}{\end{equation}}
\newcommand{\bea}{\begin{eqnarray}}
\newcommand{\eea}{\end{eqnarray}}

\newcommand{\Span}{\operatorname{span}}

\newcommand{\rank}{\operatorname{rank}}
\newcommand{\kket}[1]{| #1 \rangle}
\newcommand{\bbra}[1]{\langle #1 |}
\newcommand{\braket}[2]{\langle #1 | #2 \rangle}
\newcommand{\ketbra}[2]{| #1 \rangle \langle #2 |}

\def\real{{\mathbb{R}}}

\def\complex{{\mathbb{C}}}

\def\proba{{\rm I\kern -.18em P}}

\newcommand{\eg}{e.g.}
\newcommand{\ie}{i.e.}

\newcommand{\cv}{{\bf{c}}}
\newcommand{\ev}{{\bf{e}}}

\newcommand{\uv}{{\bf{u}}}

\newcommand{\sv}{{\bf{s}}}

\newcommand{\Cc}{{\cal C}}

\newcommand{\Hh}{{\cal H}}

\newcommand{\Mm}{{\cal M}}

\newcommand{\Ss}{{\cal S}}

\newcommand{\Vv}{{\cal V}}

\newcommand{\Aclass}{{A\text{-cl}}}

\newcommand{\opt}{{\rm{opt}}}
\newcommand{\mmax}{m_{\rm max}}

\begin{document}

\title {Geometric quantum discord with Bures distance: the qubit case}
\author{D. Spehner}
\email{Dominique.Spehner@ujf-grenoble.fr}
\affiliation{Universit\'e Grenoble 1 and CNRS, Institut Fourier  UMR5582,\\
B.P. 74, 38402 Saint Martin d'H\`eres, France}
\affiliation{Universit\'e Grenoble 1 and CNRS, Laboratoire de Physique et Mod\'elisation des
Milieux Condens\'es UMR5493, \\
B.P. 166, 38042 Grenoble, France}
\author{M. Orszag}
\affiliation{Pontificia Universidad Cat\'olica, 
Facultad de f\'{\i}sica, \\
Casilla 306, Santiago 22, Chile}  
\date{\today}


\begin{abstract}
The minimal Bures distance of a quantum state of a bipartite system $AB$
to the set of classical states for subsystem $A$ defines a geometric measure of
 quantum discord. When $A$ is a qubit, we show  that this 
geometric quantum discord is given in terms of the eigenvalues of a 
$(2 n_B) \times (2 n_B)$ hermitian matrix, 
$n_B$ being the Hilbert space dimension of the other subsystem $B$.
As a first application, we calculate the geometric discord for the output state of the DQC1 algorithm.
We find that it takes its highest value when the  unitary matrix from which the algorithm computes the trace
has its eigenvalues uniformly distributed on the unit circle modulo a symmetry with respect to the origin. 
As a second application,  we derive an explicit formula for  
 the geometric discord of a two-qubit state $\rho$ with maximally mixed marginals and compare it 
with other measures of quantum correlations.
We also determine the closest classical states to $\rho$.
\end{abstract}

\maketitle

\section{Introduction} \label{sec-intro}

In order to understand the origin of
quantum speedups in quantum algorithms and to analyze the specificities of quantum communication protocoles, 
it is of prime importance to identify the resources needed to 
have a quantum advantage~\cite{Nielsen}.
Despite substantial progress in the last decades~\cite{Horedecki_review,Modi_review}, a complete
 characterization of  quantum correlations (QCs) in composite quantum systems has not yet emerged,
even for bipartite systems. Furthermore,  
the role played by these correlations in information processing tasks remains to a large extent unclear.
The quantum discord (QD) is one measure of QCs in bipartite systems~\cite{Ollivier01,Henderson01}.
It coincides with the entanglement of formation~\cite{Bennett96} for pure states, 
but for mixed states it quantifies 
QCs  that may be present even in separable states. The discord 
is believed to be a better indicator  than entanglement of the ``degree of quantumness'' of mixed states.
Moreover, it has been suggested~\cite{Datta08,Lanyon08,Passante11,Fanchini11} that it could
capture the relevant quantum resource  in  Knill and Laflamme's algorithm of
deterministic quantum computation with one qubit (DQC1)~\cite{Laflamme98}.
This algorithm allows to compute efficiently the trace of a $2^n \times 2^n$ unitary matrix  $U_n$.
It involves a polarized control qubit, which remains unentangled with  $n$ unpolarized target qubits
at all stages of the computation. The amount of entanglement for any bipartition of the 
$(n+1)$ qubits is bounded independently of $n$~\cite{Datta05}. 
However, a non-vanishing QD between the control and the $n$ target qubits appears in the output 
state~\cite{Datta08,Lanyon08,Passante11,Fanchini11}, save
for some particular unitaries $U_n$~\cite{Lanyon08,Dakic10}.    

In a recent article~\cite{companion_paper}, we have proposed to use  the Bures distance to 
the set of zero-discord states as a geometrical analogue of the QD. 
This geometric QD (GQD)  does not suffer the drawbacks of a similar measure introduced
by Daki\'c, Vedral, and Brukner~\cite{Dakic10}, which makes use of the Hilbert-Schmidt distance.
 In particular,  for pure states the Bures-GQD  
is equal to the geometric measure of entanglement defined as the Bures distance to the 
set of separable states~\cite{Wei03,Streltsov10}.
The main advantage of the geometric approach is that, in addition to quantifying the degree of
quantumness of a given state $\rho$, one can look for the closest zero-discord
states to $\rho$. This may help to understand decoherence processes and  peculiar features 
of QCs during dynamical evolutions, such as the sudden transitions discussed in~\cite{Maziero09,Mazzola10}. 
The main result of Ref.~\cite{companion_paper} is that the Bures-GQD of a mixed state $\rho$  
is equal to the maximal success probability in ambiguous quantum state discrimination 
of a family of states $\rho_i$ depending on $\rho$. Moreover, the closest zero-discord states to $\rho$ 
are given in terms
of the optimal von Neumann measurement in this discrimination task. 

In this paper, we apply the aforementioned results~\cite{companion_paper} in order to calculate 
the Bures-GQD when one subsystem of the bipartite system is a qubit.
In particular, we determine the GQD  between the control qubit and the $n$ target
qubits of the output state in the DQC1 algorithm. 
We show that for any unitary matrix $U_n$, the discord is bounded from above  
by its value obtained by choosing random unitaries distributed according to the Haar measure. 
We also derive an explicit formula for  states 
$\rho$ of two qubits with maximally mixed marginals
and  determine the closest  zero-discord states to $\rho$. 
We compare our results to those obtained by using the Hilbert-Schmidt distance and the relative entropy.

The layout of the paper is as follows. We recall in Sec.~\ref{sec-one_qubit_general} the problem
of ambiguous quantum state discrimination (QSD), the definitions of the quantum discords, and the link between 
the geometric discord and QSD. 
In Sec.~\ref{sec-2_qubit_general}, we show how to solve the 
QSD task  and determine the GQD when the measured subsystem is a qubit.
We calculate in Sec.~\ref{sec-DQC1} the GQD for the output  state 
in the DQC1 model.
Sec.~\ref{sec-GDQ_Luo_states} is devoted to the case of two qubits. 
We compare our results for the GQD 
with the other quantum discords and find  the closest zero-discord states
to a Bell-diagonal state.    
The last section~\ref{sec-Conclusion} contains conclusive remarks.
Some technical details are presented in the appendix.

\section{Quantum state discrimination and the geometric quantum discord} 
\label{sec-one_qubit_general}
\subsection{Ambiguous quantum state discrimination} \label{sec-QSD}

The objective of quantum state discrimination (QSD) consists in distinguishing
 states taken randomly from a known ensemble of states~\cite{Helstrom,Bergou_review}.
If these states are  non-orthogonal, any measurement devised to distinguish them cannot succeed
to identify exactly which state from the ensemble has been chosen. 
The QSD task is to find the optimal 
measurements leading to the smallest probability of equivocation.
More precisely, a receiver is given a state $\rho_i$  in a finite dimensional 
Hilbert space $\Hh$, drawn from a known family
$\{ \rho_i \}_{i=1}^{n_A}$, with a prior probability $\eta_i$.
In order  to determine which state he has received, he performs a generalized measurement
and concludes that the state is $\rho_j$ when he gets the measurement outcome $j$.
The probability to find this outcome given that the state is
$\rho_i$ is $P_{j|i}=\tr ( M_j \rho_i )$, where $\{ M_j\}$ is a family of 
non-negative measurement operators satisfying 
$\sum_j  M_j = 1$ (POVM). 
In the so-called ambiguous QSD strategy, the number of 
measurement outcomes is chosen to be equal to the number $n_A$ of states in the ensemble $\{ \rho_i,\eta_i\}$.  
The maximal success probability of the receiver reads 
\begin{equation} \label{eq-max_success_proba_POVM}
P_{S}^{\,\rm{opt}} ( \{ \rho_i,\eta_i\}) =\max_{{\rm POVM}\; \{ M_i \} } \sum_{i=1}^{n_A} \eta_i \tr ( M_i \rho_i)\;.
\end{equation}
If the $\rho_i$ have rank $n_B=\dim (\Hh) /n_A$ 
 and are linearly independent, in the sense that their eigenvectors $\kket{\xi_{ij}}$
with nonzero eigenvalues 
form a linearly independent family $\{ \kket{\xi_{ij}} \}_{i=1, \ldots , n_A}^{j=1,\ldots, n_B}$
of vectors in $\Hh$, it is known~\cite{Eldar03} that the optimal POVM $\{ M_i^\opt \}_{i=1}^{n_A}$
 is a von Neumann measurement with projectors of rank $n_B$. Therefore, in that case
$P_{S}^{\,\rm{opt}} ( \{ \rho_i,\eta_i\}) =P_S^{\,\rm{opt\,v.N.}} ( \{ \rho_i,\eta_i \})$, where
\begin{equation} \label{eq-max_success_proba_von_Neumann}
P_S^{\,\rm{opt\,v.N.}} ( \{ \rho_i,\eta_i \})
= \max_{ \{ \Pi_i \} } \sum_{i=1}^{n_A} \eta_i \tr ( \Pi_i \rho_i) 
\end{equation}
is the maximum success probability over all orthogonal families $\{ \Pi_i\}_{i=1}^{n_A}$ of 
projectors of rank $n_B$ (\ie, self-adjoint operators satisfying 
$\Pi_i \Pi_j = \delta_{ij} \Pi_i$ and $\rank (\Pi_i) = n_B$). 

\subsection{Quantum discords} \label{sec-def_QD}

We briefly recall the definition of the QD of\- Ollivier and Zurek and 
Henderson and Vedral~\cite{Ollivier01,Henderson01}.
Let $AB$ be a bipartite system  in the state $\rho$. 
The mutual information of $AB$ is given by  
$I_{A:B}(\rho )= S(\rho_A) +  S(\rho_B)-S(\rho)$, 
where $S(\cdot )$ stands for the von Neumann entropy and $\rho_A = \tr_B (\rho)$ and
$\rho_B = \tr_A ( \rho)$ are the reduced states of $A$ and $B$, respectively.
The mutual information is non-negative by the  sub-additivity of $S$. It characterizes the 
total (classical and quantum) correlations in $AB$. 
The QD measures the amount of mutual information which is not accessible 
by local measurements on the subsystem $A$. It can be defined as
\begin{equation} \label{eq-def_discord}
\delta_A (\rho) 
 =  
   I_{A:B} ( \rho) - \max_{ \{ \pi_i^A \} } I_{A:B} ( \Mm_{ \{ \pi^A_i\} } (\rho) ) \, ,
\end{equation}
where the maximum is over all von Neumann measurements (\ie, orthogonal families of projectors) 
$\{ \pi_i^A\}$ on  $A$ and 
$\Mm_{\{ \pi^A_i\} } (\rho) = \sum_i \pi_i^A \otimes 1 \,\rho\,\pi_i^A \otimes 1$
is the  post-measurement state in the absence of readout.
The second term in (\ref{eq-def_discord}) represents the amount of classical correlations.
It can be shown that $\delta(\rho) \geq 0$  and $\delta_A(\sigma_{\Aclass} )=0$ if and only if 
\be \label{eq-A-classical_state}
\sigma_{\Aclass} 
=
\sum_{i=1}^{n_A}  q_i \ketbra{\alpha_i}{\alpha_i} \otimes \sigma_{B|i} \;,
\ee
where $\{ \kket{\alpha_i} \}_{i=1}^{n_A}$ is an orthonormal basis for subsystem $A$, 
$\sigma_{B|i}$ are some (arbitrary)
states of $B$ depending on the index $i$, and $q_i \geq 0$ are some probabilities.
We call {\it $A$-classical states}
the zero-discord states of the form (\ref{eq-A-classical_state})~\cite{remark1}.

The set of quantum states can be equipped with various distances.
From a quantum information perspective, it is natural to study the geometry
induced by the Bures distance~\cite{Nielsen,Bures69,Uhlmann76}
\be \label{eq-Bures_distance}
d_B ( \rho, \sigma) = \Bigl[ 2 \bigl( 1 - \sqrt{F(\rho,\sigma)} \bigr) \Bigr]^{\frac{1}{2}}\,,
\ee
where $F(\rho,\sigma) $ is the Uhlmann fidelity
\be \label{eq-fidelity}
F(\rho,\sigma) 
= \| \sqrt{\rho} \sqrt{\sigma} \|_1^2 
 = \bigl[ \tr \bigl(  [ \sqrt{\sigma} \rho \sqrt{\sigma} ]^{1/2} \bigr) \bigr]^2 
\ee
generalizing the usual pure-state  fidelity
$| \braket{\Psi}{\Phi} |^2$.
The distance $d_B$ pertains to the family of monotonous (that is, contractive with respect to 
completely positive trace-preserving maps) Riemannian distances~\cite{Petz96}. Its metric coincides  with 
the quantum Fisher information playing an important role in 
quantum metrology~\cite{Caves94}. We take this opportunity to mention
a misprint in our previous article~\cite{companion_paper}:
 although the square distance 
$d_B (\rho,\sigma)^2$
is jointly convex, this is not the case for $d_B(\rho,\sigma)$.
 
The geometric QD  (GQD) is by definition the square distance of 
$\rho$ to the set $\Cc_A$ of $A$-classical states,
\begin{equation} \label{eq-def_geo_discord}
D_A (\rho) = d_B ( \rho, \Cc_A)^2 \equiv \min_{\sigma_\Aclass \in \Cc_A} d_B ( \rho,\sigma_\Aclass )^2\;.
\end{equation} 
One can similarly define the discord $D_B(\rho)=d_B ( \rho, \Cc_B)^2$, 
where $\Cc_B$ is the set of $B$-classical states. One has in general $D_B \not= D_A$ (in fact, 
it is clear from the form~(\ref{eq-A-classical_state}) of the $A$-classical states that $\Cc_B \not= \Cc_A$).
The same holds for the discords $\delta_A \not= \delta_B$, because
the maximal mutual informations after measuring subsystems $A$ or $B$
are in general different.
It is not difficult to show~\cite{companion_paper} that for pure states 
$D_A$ and $D_B$ coincide with the geometric measure of entanglement
$E(\rho)=\min_{\sigma_{\rm sep} \in \Ss} d_B (\rho, \sigma_{\rm sep})^2$, 
where $\Ss$ denotes the convex set
of separable states. More precisely, if $\rho_\Psi = \ketbra{\Psi}{\Psi}$ then
$D_A (\rho_\Psi) = D_B (\rho_\Psi)=E(\rho_\Psi ) = 2(1 - \sqrt{\mu_{\rm max}})$,
$\mu_{\rm max}$ being the highest eigenvalue  of $(\rho_\Psi)_A$ 
(maximal Schmidt coefficient).
This equality between $D_A$, $D_B$, and $E$ comes from the 
fact that the closest separable state to a pure state  is  a pure product state~\cite{remark0}.

\subsection{Link between $D_A$ and state discrimination}
\label{sec-link_D_A_QSD}

The evaluation of the GQD (\ref{eq-def_geo_discord}) for mixed states $\rho$ turns out to be related
to an ambiguous QSD task~\cite{companion_paper}.
More precisely,  the fidelity between $\rho$ and its closest $A$-classical state is given by 
the maximum success probability (\ref{eq-max_success_proba_von_Neumann}),
\begin{equation} \label{eq-variationnal_formula_bis}
F_A (\rho) \equiv  \max_{\sigma_\Aclass \in \Cc_A} F (\rho, \sigma_\Aclass)
=\max_{\{ \kket{\alpha_i} \} } P_S^{\,\rm{opt\,v.N.}} ( \{ \rho_i,\eta_i \})\,.
\end{equation}
In the right-hand side, the maximum  
is over all orthonormal basis $\{ \kket{\alpha_i} \}_{i=1}^{n_A}$ of $A$
and $\{ \rho_i, \eta_i \}_{i=1}^{n_A}$ is an ensemble of states depending 
on $\{ \kket{\alpha_i} \}$ and  $\rho$  defined by
\begin{equation} \label{eq-state_Q_discrimination}
\eta_i = \bbra{\alpha_i}  \rho_A \kket{\alpha_i} \; , \; 
\rho_i = \eta_i^{-1} \sqrt{\rho} \ketbra{\alpha_i}{\alpha_i} \otimes 1 \sqrt{\rho}
\end{equation}
(if $\eta_i = 0$ then $\rho_i$ is not defined but does not contribute to 
the success probability).
The number  of states $\rho_i$ is equal to the dimension $n_A$
of the Hilbert space of $A$.
It is easy to see that if $\rho>0$ then all the $\rho_i$ have ranks equal to the dimension $n_B$ 
of the space of $B$. Furthermore, they are 
linearly independent. Thus $P_S^{\,\rm{opt\,v.N.}}$ can be replaced by 
$P_S^{\,\rm{opt}}$ in Eq.(\ref{eq-variationnal_formula_bis}).

Note that $\{ \rho_i,\eta_i\}$ defines a convex decomposition of $\rho$,  
$\rho = \sum_i \eta_i \rho_i$. 
A remarkable property of this decomposition is that  
the associated square-root measurement operators  
$M_i  = \eta_i \rho^{-1/2}  \rho_i \rho^{-1/2}$,
which are known to be optimum in the ambiguous QSD of symmetric  ensembles~\cite{Eldar04,Chou03}, 
coincide with the projectors
$\Pi_i =\ketbra{\alpha_i}{\alpha_i} \otimes 1$.
By bounding from below $P_S^{\,\rm{opt\,v.N.}} ( \{ \rho_i,\eta_i \})$ by
the success probability corresponding to $\Pi_i$, we obtain
\begin{equation} \label{eq-inequality_square_root_meas}
F_A (\rho) \geq \max_{\{ \kket{\alpha_i} \} } \sum_{i=1}^{n_A} 
 \tr_B \bigl[ \bbra{\alpha_i} \sqrt{\rho} \kket{\alpha_i}^2 \bigr]\;.
\end{equation}

Let us denote by $\{ \kket{\alpha_i^{\rm{opt}}} \}$ and 
$\{ \Pi_i^{\rm{opt}} \}$ the  
basis  and projective measurement(s) maximizing  
$P_S^{\,\rm{opt\,v.N.}} ( \{ \rho_i,\eta_i \})$ in (\ref{eq-variationnal_formula_bis}).
Then  the closest $A$-classical state(s) to $\rho$ is (are)~\cite{companion_paper}
\begin{equation} \label{eq-again_I_was_stupid}
\sigma_\rho = \frac{1}{F_A(\rho)} 
\sum_{i=1}^{n_A} \ketbra{\alpha_i^{\rm{opt}}}{\alpha_i^{\rm{opt}}} 
\otimes \bbra{\alpha_i^{\rm{opt}}} \sqrt{\rho}\, \Pi_i^{\rm{opt}} \sqrt{\rho} \kket{\alpha_i^{\rm{opt}}} \,. 
\end{equation}

Finding analytically the optimal success
probability and optimal measurement(s) in ambiguous QSD with $n_A>2$ 
states is an open problem, excepted for symmetric ensembles~\cite{Bergou_review,Eldar04,Chou03}.
A necessary and sufficient condition for a POVM to be optimum is due to Helstrom~\cite{Helstrom}.    
Various  bounds on $P^\opt_S$ have been derived in the literature (see~\cite{Qiu10} and references
therein). Moreover,  efficient methods are available to solve this problem
numerically~\cite{Helstrom82,Jezek12}.
We focus in the next section on the simpler case $n_A=2$, for which
an analytic solution is well-known.

\section{Geometric discord for a $(2,n_B)$ bipartite system} \label{sec-2_qubit_general}

If the subsystem $A$ is a qubit, the ensemble $\{ \rho_i, \eta_i \}$ contains only $n_A=2$ states
and the QSD task can be easily handled~\cite{Helstrom,Bergou_review}. 
In our case, we must maximize the success probability over all von Neumann measurements 
given by orthogonal projectors $\Pi_0$ and $\Pi_1$ having the same rank $n_B$. 
One starts by writing the projector $\Pi_1$ as $1-\Pi_0$ in the expression of the success probability,
\begin{eqnarray}  \label{eq-success_proba}
\nn
P_{S}^{\{ \Pi_i \} } ( \{ \rho_i,\eta_i\}) 
& = & \eta_0 \tr ( \Pi_0 \rho_0 ) + \eta_1 \tr ( (1-\Pi_0) \rho_1 )
\\
& = &  
\frac{1}{2} \bigl( 1 - \tr  \Lambda   \bigr)+ \tr ( \Pi_0 \Lambda )  
\end{eqnarray}
with $\Lambda = \eta_0 \rho_0 - \eta_1 \rho_1$.
Thanks  to the min-max principle~\cite{Bhatia}, the maximum of $\tr ( \Pi_0 \Lambda)$ 
 over all projectors $\Pi_0$ of rank $n_B$ 
is equal to the sum of the $n_B$ highest eigenvalues $\lambda_1  \geq \cdots \geq \lambda_{n_B}$
of the hermitian matrix $\Lambda$, and
the optimal projector $\Pi_0^{\rm{opt}}$ is the spectral projector associated to these highest eigenvalues.
One has 
\begin{eqnarray}  \label{eq-opt_success_proba}
\nn
P_S^{\,\rm opt\,v.N.}( \{ \rho_i,\eta_i\}) 
& = & \max_{ \{ \Pi_i\} } P_{S}^{\{ \Pi_i \}} ( \{ \rho_i,\eta_i\})
\\
& = & 
\frac{1}{2}\bigl(  1- \tr  \Lambda \bigr) + 
\sum_{l=1}^{n_B} \lambda_l \;.
\end{eqnarray}
For the states $\rho_i$ and probabilities $\eta_i$ defined by Eq.(\ref{eq-state_Q_discrimination}),   
$\Lambda=\sqrt{\rho} \,( \ketbra{\alpha_0}{\alpha_0} - \ketbra{\alpha_1}{\alpha_1} ) \otimes 1\, \sqrt{\rho}$,
where $\{ \kket{\alpha_i}\}_{i=0}^1$ is an orthonormal basis of $\complex^2$. 
For any such basis the operator inside the parenthesis in the last formula
is equal to  $\sigma_{\uv} \equiv \sum_{m=1}^3 u_m \sigma_m$ for some unit vector $\uv \in \real^3$
(here $\sigma_1$, $\sigma_2$, and $\sigma_3$  are the Pauli matrices).
Reciprocally,  one can associate to  any unit vector $\uv \in \real^3$ the 
eigenbasis $\{ \kket{\alpha_i}\}_{i=0}^1$ of ${\sigma}_{\uv}$.
By  substituting (\ref{eq-opt_success_proba}) into (\ref{eq-variationnal_formula_bis}), we get
\begin{equation} \label{eq-fidelity_as_min_success_discrimination_qubit}
F_A (\rho) =  
\frac{1}{2} \max_{\| \uv \|=1 } \Bigl\{  1  -   \tr  \Lambda (\uv)  
+ 2 \sum_{l=1}^{n_B} \lambda_l (\uv)  \Bigr\} 
\end{equation}
where
$\lambda_l(\uv)$ are the eigenvalues in non-increasing order of the $2n_B \times 2 n_B$ hermitian matrix 
\begin{equation} \label{eq-Lambda}
 \Lambda (\uv )  =  \sqrt{\rho} \, {\sigma}_{\uv} \otimes 1 \, \sqrt{\rho} \;.
\end{equation}
Note that $-\rho \leq \Lambda (\uv ) \leq \rho$, so that 
$\sum_{l=1}^{n_B} \lambda_l (\uv) \leq \sum_{l=1}^{n_B} p_l$, $p_l$ being the eigenvalues of $\rho$
in non-increasing order. 
If $\rho>0$ then $\Lambda (\uv)$ has $n_B$ positive and $n_B$  
negative eigenvalues. In such a case
(\ref{eq-opt_success_proba}) reduces to the well known expression
$P_S^{\,\rm opt\,v.N.} ( \{ \rho_i,\eta_i\})= P_S^{\,\rm opt} ( \{ \rho_i,\eta_i\}) 
 = (1+ \tr | \Lambda (\uv) |)/2$~\cite{Bergou_review}.

It is shown in~\cite{companion_paper} that the minimal value that $F_A(\rho)$  can take
is equal to $1/n_A$ when $n_A \leq n_B$~\cite{remark2}.
If $A$ is a qubit, the maximal value of the GQD is thus $D_A=2 - \sqrt{2}$,
see~(\ref{eq-Bures_distance}).
It is convenient to work with a normalized discord, given in terms of  $F_A(\rho)$  by
\begin{equation} \label{eq-normalizedGDQ} 
\widetilde{D}_A ( \rho) = \frac{D_A ( \rho)}{2 - \sqrt{2}}  = \frac{1 - \sqrt{ F_A (\rho )}}{1-1/\sqrt{2}} \;.
\end{equation}
%

\section{Geometric discord in the  DQC1 model} \label{sec-DQC1}

In this section, we determine the Bures-GQD for the output state of the DQC1 algorithm.
We obtain an analytic expression of the discord for all unitary matrices $U_n$
from which the algorithm computes the trace.

The DQC1 model~\cite{Laflamme98} consists in a control qubit, labelled by the index $0$, coupled to
$n$ target qubits $1,2,\ldots, n$. The control qubit is initially in a mixture of the  
 standard basis vectors $\kket{0}$ and $\kket{1}$
with populations $(1\pm \alpha)/2$ differing from one half.
The target qubits are in the completely mixed state $2^{-n}\,1_n$, where $1_n$ stands for 
the identity operator on the $2^n$-dimensional space $\Hh_B$ of qubits $1,\ldots,n$. This state
approximates well the thermal state of the nuclear spins in liquid-state 
nuclear magnetic resonance (NMR) experiments at room temperature. The initial state of the $(n+1)$ qubits reads
\begin{equation}
\rho_{n+1,{\rm in}}^{\alpha} =
\frac{1}{2^{n+1}} ( 1 + \alpha \,\sigma_3 ) \otimes 1_n 
\end{equation}
with $\alpha \in [-1,1]$, $\alpha \not=0$. 
This state is transformed by the following circuit composed
of a Hadamard gate $H$ and a unitary gate $U_n$ acting on $\Hh_B$ controlled
by the qubit $0$:
\begin{equation}
\Qcircuit @C=1.5em @R=1em {
\lstick{\frac{1}{2} ( 1 + \alpha\, \sigma_3)} & \gate{H}  & \ctrl{1} & \meter & \\
\lstick{\frac{1}{2^n} 1_n} & \qw      & \gate{U_n} & \qw &
\;}
\end{equation}
The output state is
\begin{eqnarray} \label{eq-output_DQC1}
\nn
& & \rho_{n+1}^{U,\alpha} 
 =  
\frac{1}{2^{n+1}} 
\left( \begin{array}{cc}
 1_n &  \alpha U_n^\dagger \\[0.7em]
\alpha U_n & 1_n 
\end{array}
\right)
 = \frac{1}{2 N}
\Bigl( \ketbra{0}{0} \otimes 1_n + 
\\
& & \hspace*{5mm} \ketbra{1}{1} \otimes 1_n + \alpha \ketbra{0}{1} \otimes U_n^\dagger
+ \alpha \ketbra{1}{0} \otimes U_n \Bigr) 
\end{eqnarray} 
with $N = 2^n$. Hence the reduced state $\tr_{1,\ldots,n} (\rho_{n+1}^{U,\alpha} )$ of the qubit $0$
contains some information about the normalized trace $z_n = \tr (U_n)/N$.
No efficient classical algorithm to compute $z_n$ is known.
 The DQC1 quantum algorithm provides accurate approximations of 
$\langle \sigma_1 \otimes 1_n \rangle_{\rm out} = \alpha\, \re z_n$ and 
$\langle \sigma_2 \otimes 1_n \rangle_{\rm out} = \alpha\, \im z_n$ after 
sufficiently many runs of the measurement of the spin 
$\sigma_1$ and $\sigma_2$ on the qubit $0$.
The number of runs is independent of $n$ and scales logarithmically with 
the error probability. In that sense, the DQC1 algorithm is exponentially more efficient than all known classical 
algorithms for estimating the {\it normalized} trace of $U_n$~\cite{Datta05}.
Moreover, it works whatever the value of $\alpha$ provided that $\alpha \not= 0$, hence
demonstrating the ``power of even the tiniest fraction of a qubit''~\cite{Datta05}.

The control qubit is unentangled with the target qubits, as can be shown from the relation
\begin{equation} \label{eq-output_state_DQC1bis}
\rho_{n+1}^{U,\alpha} =
\frac{1}{2N} \sum_{k=1}^{N} \bigl( \ketbra{\varphi_k}{\varphi_k} + \ketbra{\chi_k}{\chi_k} \bigr) \otimes \ketbra{u_k}{u_k}\,,
\end{equation}
where $\kket{u_k}$ are the eigenvectors of $U_n$ with eigenvalues $e^{\I \omega_k}$, 
$\kket{\varphi_k} = \cos \delta \kket{0} + e^{\I \omega_k} \sin \delta \kket{1}$, and
$\kket{\chi_k} = \sin \delta \kket{0} + e^{\I \omega_k} \cos \delta \kket{1}$,
with $\sin(2 \delta ) = \alpha$.
For other bipartitions of the $(n+1)$ qubits, \eg, 
putting  together in one subsystem the control qubit and half of the target qubits, 
the entanglement does not vanish in general but it is bounded 
uniformly in $n$~\cite{Datta05}. For large system sizes, 
the total amount of bipartite entanglement is thus a negligible 
fraction of the maximal entanglement possible.

In what follows, we split the $(n+1)$ qubits into the subsystem $A$ containing the qubit $0$ and the subsystem $B$
containing the qubits $1, \ldots, n$. It is clear from (\ref{eq-output_state_DQC1bis})
that the output state is $B$-classical, hence it has a vanishing discord $\delta_B$.
It was shown in~\cite{Datta08} that this state has  generically a non-vanishing discord
$\delta_A$. The term  ``generically'' refers here to a 
random choice of the unitary $U_n$ with the Haar distribution.  
 This presence of a nonzero discord  
 has been demonstrated experimentally in optical~\cite{Lanyon08} and 
liquid-state NMR~\cite{Passante11} implementations of DQC1.

The Bures-GQD of the output state (\ref{eq-output_DQC1}) can be easily
determined  with the help of (\ref{eq-fidelity_as_min_success_discrimination_qubit}).
The eigenvalues of $\rho_{n+1}^{U,\alpha}$ are $(1\pm \alpha)/(2N)$ and the two corresponding 
eigenspaces are spanned by the $N$ vectors $\kket{0} \ket{\underline{j}} \pm \kket{1} U_n \ket{\underline{j}}$,
where $\kket{\underline{j}}$, $\underline{j} \in \{ 0, 1 \}^n$,
 are the standard basis vectors of $\Hh_B$.
This yields the square root 
\begin{eqnarray}
\nn
\sqrt{ \rho_{n+1}^{U,\alpha}} 
& = & 
\frac{1}{\sqrt{2N}} V 
\left( \begin{array}{cc}
 \sqrt{1 + \alpha}\,\, 1_n &  0 \\[0.7em]
 0 & \sqrt{1-\alpha}\,\, 1_n 
\end{array}
\right) V^\dagger 
\\[0.7em]
V 
& = & 
\frac{1}{\sqrt{2}} \left( \begin{array}{cc}
 1_n &  1_n \\[0.7em]
 U_n & -U_n 
\end{array}
\right)\;.
\end{eqnarray} 
Substituting this expression into (\ref{eq-Lambda}) and introducing the angles $\theta,\phi$ such that 
$\uv=  (\sin \theta \cos \phi , \sin \theta \sin \phi, \cos \theta)$, one finds 
\begin{eqnarray} \label{eq-Lambda_DQC1}
\nn
\Lambda(\uv) 
& = &  
  \frac{1}{2N}\, V 
  \left( 
\begin{array}{c}
  (1+\alpha) \sin \theta \, \re ( U_n^\phi )  \\[0.7em]
 \sqrt{1-\alpha^2} (\cos \theta+\I \sin \theta \, \im ( U_n^\phi) )  
\end{array}
\right.
\\[0.9em] 
& & 
\hspace*{2mm} \left. 
\begin{array}{c}
  \sqrt{1-\alpha^2} (\cos \theta-\I \sin \theta \, \im ( U_n^\phi )) \\[0.7em]
- (1-\alpha)   \sin \theta\, \re ( U_n^\phi ) 
\end{array}
\right)
V^\dagger
\end{eqnarray}
with $U_n^\phi = e^{-\I \phi} U_n$. Here, $\re (O)=(O + O^\dagger)/2$ and $\im (O)=(O-O^\dagger)/2\I$ 
denote the real and imaginary parts  of the operator $O$.
By diagonalizing each of the four blocks of the matrix appearing
between $V$ and $V^\dagger$
in the right-hand side of (\ref{eq-Lambda_DQC1}), one sees that its eigenvalues
$\lambda_{k,\pm} (\uv)$ are the eigenvalues of $k$ distinct $2\times 2$  matrices. This  yields
\begin{eqnarray}
\lambda_{k,\pm} (\uv) 
& = &  
\frac{1}{2N} \biggl( 
 \alpha \sin \theta \cos (\omega_k^\phi )
\\
\nn
& &  
  \pm \sqrt{1 - \alpha^2 + \alpha^2 \sin^2 \theta \cos^2 ( \omega_k^\phi ) }
  \biggr)
\end{eqnarray}
for $k=1,\ldots ,N$,
where $\omega_k^\phi = \omega_k - \phi$ are the eigenphases of $U_n^\phi$.
One has clearly $\pm \lambda_{k,\pm} (\uv) \geq 0$.
 Writing the maximum over $\uv$ in (\ref{eq-fidelity_as_min_success_discrimination_qubit})
as a maximum over $\theta$ and $\phi$ and noting that the maximum over 
$\theta$ is reached for $\sin^2 \theta=1$, we get
\begin{equation}  \label{eq-max_fidelity_DQC1}
F_A [ \rho_{n+1}^{U,\alpha} ] 
 = \frac{1}{2} \max_{\phi} \biggr\{
  1 + \frac{1}{N} \sum_{k=1}^N \sqrt{1- \alpha^2 \sin^2 (\omega_k-\phi)}
\biggr\} \,.
\end{equation}
Let us point out that the maximal fidelity (\ref{eq-max_fidelity_DQC1}) to the $A$-classical states
is a decreasing function of $\alpha^2$. Hence
the geometric discord $D_A (\rho_{n+1}^{U,\alpha} )$ increases with the initial purity $(1+\alpha^2)/2$
of the control qubit.

For large system sizes $n$, the sum over $k$ in (\ref{eq-max_fidelity_DQC1}) can be replaced by
an integral over the smooth normalized spectral density $n(\omega)$ of the eigenphases $\omega_k$ of $U_n$,
\begin{eqnarray}  \label{eq-max_fidelity_DQC1_n_infinity}
\nn
F_A [ \rho_{n+1}^{U,\alpha} ] 
&  = & \frac{1}{2} \max_{\phi} \Bigr\{
   1  + \int_0^{\frac{\pi}{2}} \D \omega\,  \sqrt{1- \alpha^2 \sin^2 \omega}
\\
& & \times \bigl( n_S ( \omega + \phi) + n_S (\pi - \omega + \phi) \bigr)
\Bigr\}
\end{eqnarray}
with $n_S ( \omega) = n (\omega) + n(\omega +\pi)$.
Let us show that the smallest fidelity  is achieved for 
unitary operators $U_n$ having a constant symmetrized spectral density 
$n_S (\omega)=1/\pi$.
Actually, by substituting $\max_\phi$ 
by $\int_0^{2\pi} \D \phi /( 2\pi)$ in (\ref{eq-max_fidelity_DQC1_n_infinity}) and using 
$\int_0^{2\pi} \D \phi \,n_S (\phi) = 2$, one gets the bound
\begin{eqnarray}  \label{eq-bound_max_fidelity_DQC1}
\nn
F_A [ \rho_{n+1}^{U,\alpha} ] 
& \geq &  F_A [ \rho_{n+1}^{U_{\rm unif},\alpha} ] 
\\
& = & 
\frac{1}{2} \Bigr(
  1 + \frac{2}{\pi} \int_0^{\frac{\pi}{2}} \D \omega\, \sqrt{1- \alpha^2 \sin^2 \omega}
\Bigr) \,.
\end{eqnarray}
The inequality 
is an equality if and only if the integral in (\ref{eq-max_fidelity_DQC1_n_infinity})
is independent of $\phi$. Expanding $n_S$ as a Fourier 
series, $n_S(\omega)= \sum_p a_p e^{2 \I p\omega}$, this is equivalent to 
\begin{equation} \label{eq-I_am_not_sure_for_alpha_not1}
\sum_{p=-\infty}^\infty p a_p e^{2\I p \phi}  \int_0^{\frac{\pi}{2}} \D \omega\,\sqrt{1- \alpha^2 \sin^2 \omega}
\,\cos(2 p \omega) 
 =0
\end{equation}
for any $\phi \in [0,2\pi[$. 
For $\alpha=1$, one easily sees that the integrals in (\ref{eq-I_am_not_sure_for_alpha_not1}) do not vanish
for any $p$.
Therefore $a_p=0$ when $p \not=0$ and  $n_S (\omega)$ is constant.
Reciprocally, if $n_S(\omega)$ is constant then for any $\alpha \in [-1,1]$
the inequality in (\ref{eq-bound_max_fidelity_DQC1}) is an equality.

Let us emphasize that the bound (\ref{eq-bound_max_fidelity_DQC1}) is satisfied for all $n \geq 1$  
(in fact, it can be obtained by replacing  as before 
the maximum over $\phi$ by an integral, but in Eq.(\ref{eq-max_fidelity_DQC1}) instead 
of (\ref{eq-max_fidelity_DQC1_n_infinity})). 
As a consequence, whatever the unitary matrix $U_n$ and the system size $n$, 
the GQD is always smaller or equal to the GQD for an infinite unitary matrix $U_{\rm unif}$ with constant 
symmetrized spectral density $n_S$. Such matrices
have equidistributed eigenvalues on the unit circle
modulo a symmetry with respect to the origin.
This highest possible discord is achieved in particular for random unitaries  distributed according to 
the Haar measure on the unitary group $U(N)$, since then $n (\omega) = 1/(2\pi)$ almost surely
in the large $n$ limit~\cite{Datta08}. 

\begin{figure}
\centering
\includegraphics[width=0.4\textwidth]{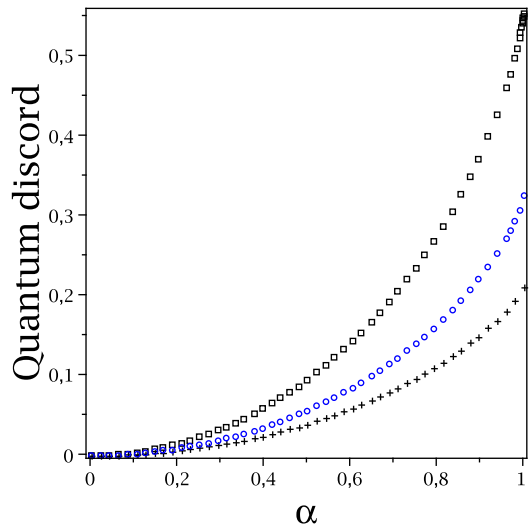}
\centering
\caption{\label{fig1}
(Color online) QDs of the output state in  
the DQC1 algorithm as a function of the purity parameter $\alpha$.
From top to bottom:
discord $\delta_A$ for random unitaries $U_n$ distributed according to
 the Haar measure in the limit $n \rightarrow \infty$ (black squares);
normalized Bures-GQD $\widetilde{D}_A$ for the same choice of unitaries (blue circles);
normalized Bures-GDQ for the rotation matrices (\ref{eq-def_unitary_rot})
 (black crosses).
}
\end{figure}

The normalized GQD (\ref{eq-normalizedGDQ}) of the output state  (\ref{eq-output_DQC1}) 
is shown in Fig.~\ref{fig1} as a function
of $\alpha$
for the optimal unitary $U_{\rm unif}$ with constant $n_S(\omega)$
and for the rotation operator
\begin{equation} \label{eq-def_unitary_rot}
U_n = \exp \biggl( \frac{\I \pi}{2 \sqrt{n}} (J_z)_n \biggr) \;\; ,\;\;n \gg 1\;,
\end{equation}
where $(J_z)_n=\sum_{i=1}^n \sigma_3^{(i)}/2$ is the total angular momentum of the $n$ spins
in the $z$ direction. In the second case, the fidelity is determined from 
(\ref{eq-max_fidelity_DQC1}) by using the non-periodic spectral density 
$n(\omega) = (2/\pi)^{3/2} \exp ( -8 \,\omega^2/\pi^2 )$
and noting that the maximum is reached for $\phi=0$. This Gaussian density is obtained by using the fact 
that the eigenphases of $U_n$, 
 $\omega_k = \pi (k-n/2)/(2 \sqrt{n})$ with $k=0,\ldots, n$,  have multiplicities 
$d_k = \frac{n!}{k! (n-k)!}$ satisfying
$2^{-n} d_k \sim \sqrt{\frac{2}{\pi n}} \exp [ - 2( k - \frac{n}{2} )^2/n ]$
in the limit $n \rightarrow \infty$
[note that the scaling like $1/\sqrt{n}$ of the rotation angle in (\ref{eq-def_unitary_rot}) 
is dictated by the wish to obtain a nontrivial $n(\omega)$]. 
We also compare in Fig.~\ref{fig1} the geometric discord with the quantum discord 
(\ref{eq-def_discord}), using the result
of Ref.~\cite{Datta08} applying to random unitaries with the Haar distribution.
We observe that 
$\delta_A (\rho_{n+1}^{U,\alpha}) \geq \widetilde{D}_A (\rho_{n+1}^{U,\alpha})$.

\begin{figure}
\centering
\includegraphics[width=0.39\textwidth]{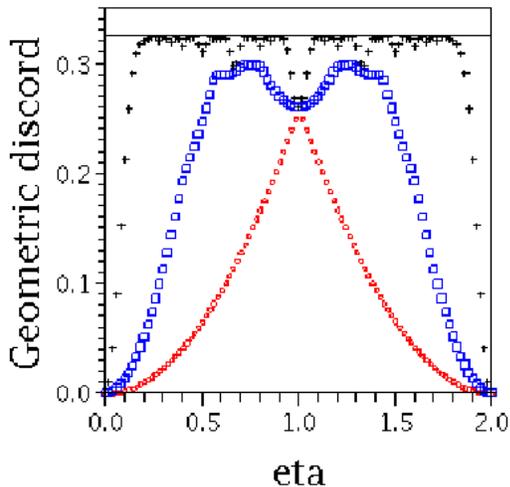}
\centering
\caption{\label{fig2}
(Color online) Normalized discord $\widetilde{D}_A$    of the output state in  
the DQC1 algorithm for  $U_n(\eta)=\exp [ \I \frac{\pi}{2} \eta (J_z)_n ]$
as a function of $\eta$. The control qubit is 
initially in a pure state ($\alpha=1$). 
From top to bottom: $n=100$ (black crosses), $n=5$ (blue squares), and $n=1$ (red diamonds).
All discords are periodic with period $2$ (not shown) and symmetric with respect to $\eta=1$. 
The horizontal line gives the maximal value $\widetilde{D}_A =(1-
\sqrt{1/2+1/\pi})/(1-1/\sqrt{2})$ obtained for random unitaries
as in Fig.~\ref{fig1}.
 }
\end{figure}

Finally, let us determine the unitaries $U_n$ leading to the smallest discord.
The fidelity (\ref{eq-max_fidelity_DQC1}) takes its maximal value $1$ when 
 $e^{\I \omega_k} \in \{ e^{\I \phi} , - e^{\I \phi}\}$ $\forall\;k=1,\ldots, N$,
for a fixed angle $\phi \in [0,2\pi[$.
Hence $D_A (\rho_{n+1}^{U,\alpha} )=0$ if and only if $U_n = e^{\I  \phi} Q$ with $\phi \in [0,2\pi[$,
$Q=Q^\dagger$, and $Q^2=Q$.
Fig.~\ref{fig2} displays the GQD  for finite system sizes $n$. 
In agreement with the aforementioned 
criterium, the discord vanishes for  rotation matrices $U_n$ with rotation angles equal to $0$ or $\pi$. 
Moreover, it 
is independent of $n$ when the rotation angle is equal $\pi/2$.
The existence of output states of the DQC1 model with a zero  discord and the
necessary and sufficient condition stated above have already been  discussed in Ref.~\cite{Dakic10}.
Let us notice that the particular matrices $U_n$ which have been conjectured in~\cite{Datta05} to lead
 for $\alpha=1$ to the highest possible bipartite entanglement among the $(n+1)$ qubits 
satisfy this condition. For such $U_n$ the discord between the control
qubit and the $n$ target qubits thus vanishes.

\vspace*{5mm}
 
\section{2-qubit states with maximally mixed marginals} \label{sec-GDQ_Luo_states}

This section is devoted to the determination of the geometric discord $D_A$ and 
the closest $A$-classical states for two-qubit systems.
With the aim of comparing $D_A$ with the other quantum discords, we
consider a simple convex family of two-qubit density matrices for which all discords can be easily
calculated: the states $\rho$ with maximally mixed marginals $\rho_A = \rho_B =1/2$. 
Any such state can be written up to a conjugation by a local unitary $U_A \otimes U_B$ (which leaves  
all discords unchanged) as~\cite{Horodecki96}
\begin{equation} \label{eq-state_with_max_disord_marginals}
\rho 
= \frac{1}{4} \Bigl( 1 \otimes 1 + \sum_{m=1}^3 c_m \sigma_m \otimes \sigma_m \Bigr)
\end{equation}
where the real vector $\cv = (c_1,c_2,c_3)$ belongs to the tetrahedron with vertices
 $F_\pm = (\pm 1,\mp 1,1)$ and $G_\pm =(\pm 1, \pm 1,-1)$, that is,
\begin{eqnarray} \label{eq-expression_p_m}
p_0  & = & \frac{1}{4} ( 1-c_1-c_2-c_3) \geq 0  \; , \; 
\\
\nn
p_m & = & \frac{1}{4} ( 1+c_1 + c_2 + c_3-2 c_m ) \geq 0 \; , \;
m=1,2,3\;.
\end{eqnarray}
The vertices $F_{\pm}$ and $G_\pm$ correspond to the Bell states
$\kket{\Phi^\pm} = ( \kket{00} \pm \kket{11})/\sqrt{2}$ and $\kket{\Psi^\pm} = ( \kket{01} \pm \kket{10})/\sqrt{2}$,
respectively. The non-negative numbers $p_\nu$ are the probabilities of
these four Bell states in
$\rho  = \sum_{\nu=0}^3  p_\nu \ketbra{\Psi_\nu}{\Psi_\nu}$,
with $\kket{\Psi_0} = \kket{\Psi^-}$, $\kket{\Psi_1} = \kket{\Phi^-}$, $\kket{\Psi_2}=\kket{\Phi^+}$, and
$\kket{\Psi_3}=\kket{\Psi^+}$.
The states (\ref{eq-state_with_max_disord_marginals}) 
form a $3$-parameter convex subset ${\mathcal T}$ in the $15$-parameter set ${\mathcal{E}}$
of all two-qubit states.

\subsection{Explicit formula for the GQD} \label{seq-GDQ_for_states_with_max_mixed_marg}

Let us calculate $D_A(\rho)$ for two qubits in the state (\ref{eq-state_with_max_disord_marginals}).
The matrices $\rho$ and $\sqrt{\rho}$ are given in the standard basis
$\{ \kket{00},\kket{10},\kket{01},\kket{11} \}$ by
\begin{equation}  \label{eq-Luo_density_matrix1}
\rho 
 = 
\frac{1}{4}
\left( \begin{array}{cccc}
1+c_3 & 0 & 0 & c_1-c_2 \\[2mm]
0 & 1-c_3 & c_1+c_2 & 0\\[2mm]
0 & c_1+c_2 & 1-c_3 & 0\\[2mm]
c_1-c_2 & 0 & 0 & 1+c_3
\end{array} \right)
\end{equation}
and
\begin{equation}  \label{eq-Luo_density_matrix2} 
\sqrt{\rho} 
 =    
\left( \begin{array}{cccc}
t_0+t_3 & 0 & 0 & t_1-t_2 \\[2mm]
0 & t_0-t_3 & t_1+t_2 & 0\\[2mm]
0 & t_1+t_2 & t_0-t_3 & 0\\[2mm]
t_1-t_2 & 0 & 0 & t_0+t_3
\end{array} \right) 
\end{equation}
where
\begin{eqnarray} \label{eq-definition_t_nu}
\nn
4 t_0  & = &  \sqrt{p_0} + \sqrt{p_1} + \sqrt{p_2} + \sqrt{p_3}
\\
4 t_1 &  = &   
- \sqrt{p_0} - \sqrt{p_1} + \sqrt{p_2} + \sqrt{p_3} \;,
\end{eqnarray}
with similar formulas for $t_2$ and $t_3$ obtained by permutation of the indices $1,2,3$.
The real parameters $t_\nu$ satisfy
$4 (t_0^2 + t_1^2 +t_2^2 + t_3^2 )= \tr \rho=1$. 
Let $\theta$ and $\phi$ be the angles defined by the vector
$\uv= ( \sin \theta \cos \phi, \sin \theta \sin \phi, \cos \theta)$ on the unit sphere.
In the standard basis, the matrix $\Lambda(\uv)=\sqrt{\rho} \,\sigma_\uv\otimes 1 \sqrt{\rho}$ reads  
\begin{widetext}
\begin{equation} \label{eq-Lambda_Luo_state}
\Lambda (\uv)
= \frac{1}{4}
\left( \begin{array}{cccc}
(a_3+b_3) \cos\theta  & \zeta^\ast_\phi \sin \theta  & \xi^\ast_\phi \sin \theta  & 0\\[2mm]
\zeta_\phi \sin \theta & (a_3-b_3) \cos\theta  & 0 & \xi^\ast_\phi \sin \theta \\[2mm]
\xi_\phi \sin \theta & 0 & (-a_3+b_3) \cos \theta  & \zeta^\ast_\phi \sin \theta \\[2mm]
0 & \xi_\phi \sin \theta & \zeta_\phi \sin \theta & (-a_3-b_3) \cos \theta 
\end{array} \right)
\end{equation}
with 
 $\xi_\phi = a_1 \cos \phi + \I a_2 \sin \phi$, $\zeta_\phi= b_1 \cos \phi + \I b_2 \sin \phi$, and
\begin{equation} \label{eq-b_m_and_a_m}
b_m = 8 (t_0^2 + t_m^2) -1
\; \;,\;\;
a_1= 8 (t_0 t_1 + t_2 t_3) \;\;,\;\;
a_2= 8 (t_0 t_2 + t_1 t_3) \;\;,\;\;
a_3= 8 (t_0 t_3 + t_1 t_2)\;.
\end{equation}
One finds
\begin{equation} \label{eq-b_m_as_function_of_c_m}
a_3 = 2 \bigl( - \sqrt{p_0 p_3} + \sqrt{p_1 p_2} \bigr)
\quad , \quad 
b_3 = 2 \bigl( \sqrt{p_0 p_3} + \sqrt{p_1 p_2} \bigr)\;.
\end{equation}
Similar expressions hold for  the other coefficients $a_1$, $a_2$, $b_1$, and $b_2$ by permuting 
the indices $1,2,3$.

The eigenvalues of $\Lambda (\uv)$ come in opposite pairs $(\lambda_\pm (\uv ),-\lambda_\pm (\uv))$ 
with
\begin{equation} \label{eq-eigenvalues_Lambda}
\lambda_\pm (\uv) 
 = 
  \frac{1}{4} 
\Bigl| 
 \sqrt{ b_3^2 \cos^2 \theta + | \zeta_\phi |^2 \sin^2 \theta} 
\pm \sqrt{ a_3^2 \cos^2 \theta + | \xi_\phi |^2 \sin^2 \theta}
\Bigr|
\;,
\end{equation}
in agreement with $\tr \Lambda(\uv) = \tr ( \rho_A \sigma_\uv )  = 0$. 
But  $|a_m | \leq b_m$ by (\ref{eq-b_m_as_function_of_c_m}), thus the second square root is smaller than
the first one.   
One deduces from (\ref{eq-fidelity_as_min_success_discrimination_qubit}) that
\begin{equation} \label{eq-max_fidelity_intermediate_step}
F_A(\rho) 
 =  
\frac{1}{2} 
\biggl( 1 +  
 \max_{\theta,\phi}   \sqrt{ b_3^2 \cos^2 \theta + ( b_1^2 \cos^2 \phi + b_2^2 \sin^2 \phi)  \sin^2 \theta} 
\biggr)\;.
\end{equation}
If the $b_m$ are distinct from each other, the maximum  is reached for 
$\cos^2 \theta , \cos^2 \phi \in \{ 0, 1\}$, that is, for $\uv = \pm \ev_1$, $\pm \ev_2$, or 
$\pm \ev_3$ (here $\ev_m$ are the coordinate axis vectors). 
If $b_1=b_2>b_3$, this maximum is reached for $\uv = \cos \phi\,\ev_1 + \sin \phi \,\ev_2$ 
with $\phi \in [0,2\pi[$ arbitrary. 
Thus
\begin{equation} \label{eq-max_fidelity_and_GDQ_Luo}
F_A(\rho) = \frac{1+b_{\rm max}}{2}
\quad \text{and}  \quad 
\widetilde{D}_A (\rho) = \Bigl(1-\frac{1}{\sqrt{2}} \Bigr)^{-1} \Bigl( 1 - \sqrt{\frac{1 + b_{\rm max}}{2}} \Bigr)
\end{equation}
with 
\begin{eqnarray} \label{eq-b}
\nn
b_{\rm max}  = \max_{m=1,\ldots,3} \{  b_m \} = \frac{1}{2}
\max 
\Bigl\{
& &  \sqrt{ ( 1+c_1)^2 - (c_2-c_3)^2}+  \sqrt{ ( 1-c_1)^2 - (c_2+c_3)^2} \, , \,
\\ \nn
& & 
  \sqrt{ ( 1+c_2)^2 - (c_1-c_3)^2}+  \sqrt{ ( 1-c_2)^2 - (c_1+c_3)^2} \, , \,
\\
& & 
  \sqrt{ ( 1+c_3)^2 - (c_1-c_2)^2}+  \sqrt{ ( 1-c_3)^2 - (c_1+c_2)^2} 
\Bigr\}\;.
\end{eqnarray}
\end{widetext}
We notice that if $|c_m|$ is maximum for $m=m_{\rm max}$  then this is also true for $b_m$, \ie\; the
 maximal number inside the brackets in (\ref{eq-b}) is the $(m_{\rm max})$th one. Actually, 
one can show  from (\ref{eq-b_m_as_function_of_c_m}) and (\ref{eq-expression_p_m}) that 
\begin{equation} \label{eq-order_c_m_and_b_m}
c_m^2-c_k^2 = b_m^2-b_k^2 \quad , \quad m,k=1,\ldots, 3 \;.
\end{equation}

\begin{figure}
\flushleft{\bf (a)}

\vspace*{-6mm}

\begin{center}
\includegraphics[width=0.29\textwidth]{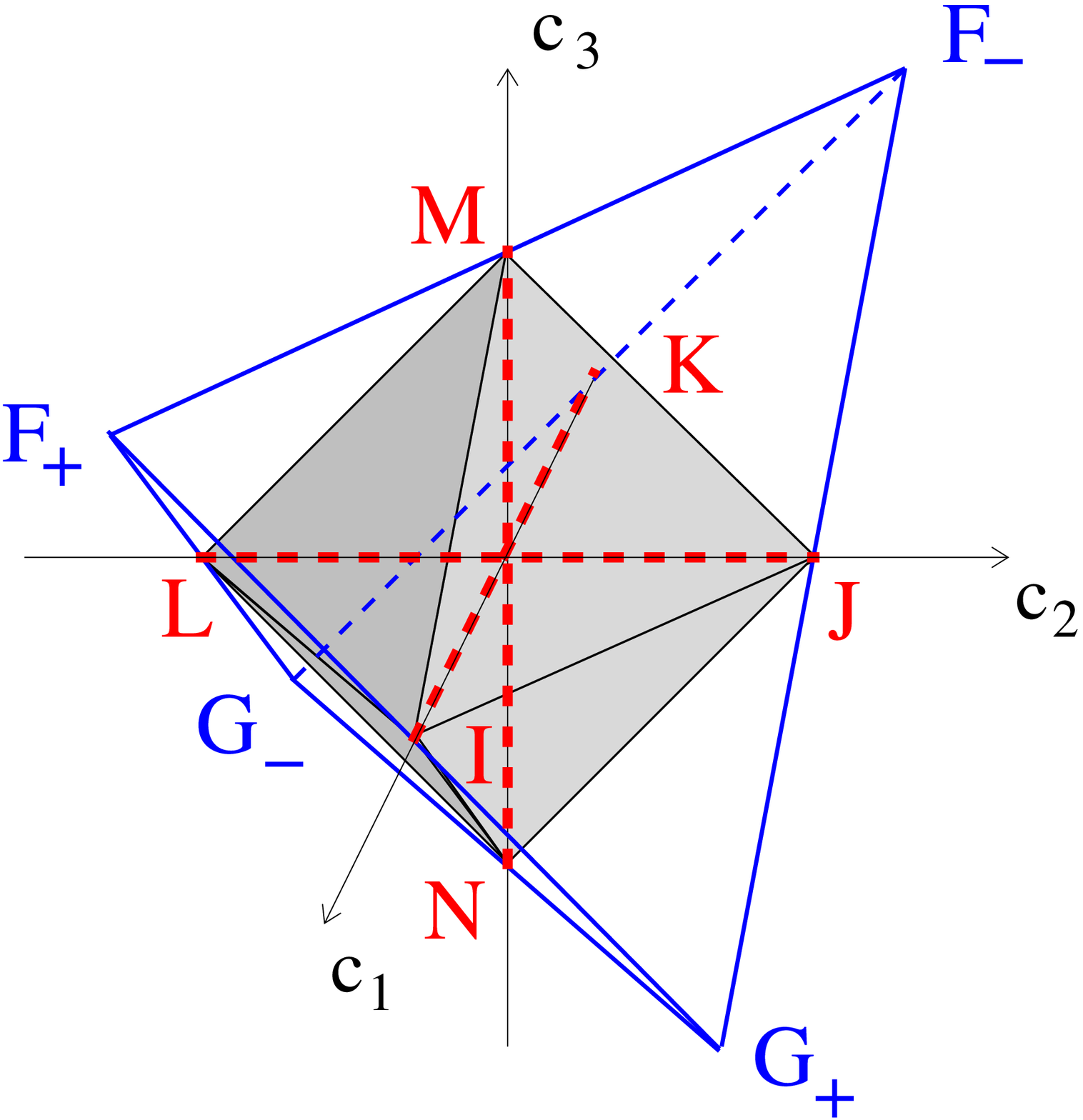}
\end{center}

\vspace*{-7mm}

\flushleft{\bf (b)}

\vspace*{-7mm}

\begin{center}
\includegraphics[width=0.23\textwidth]{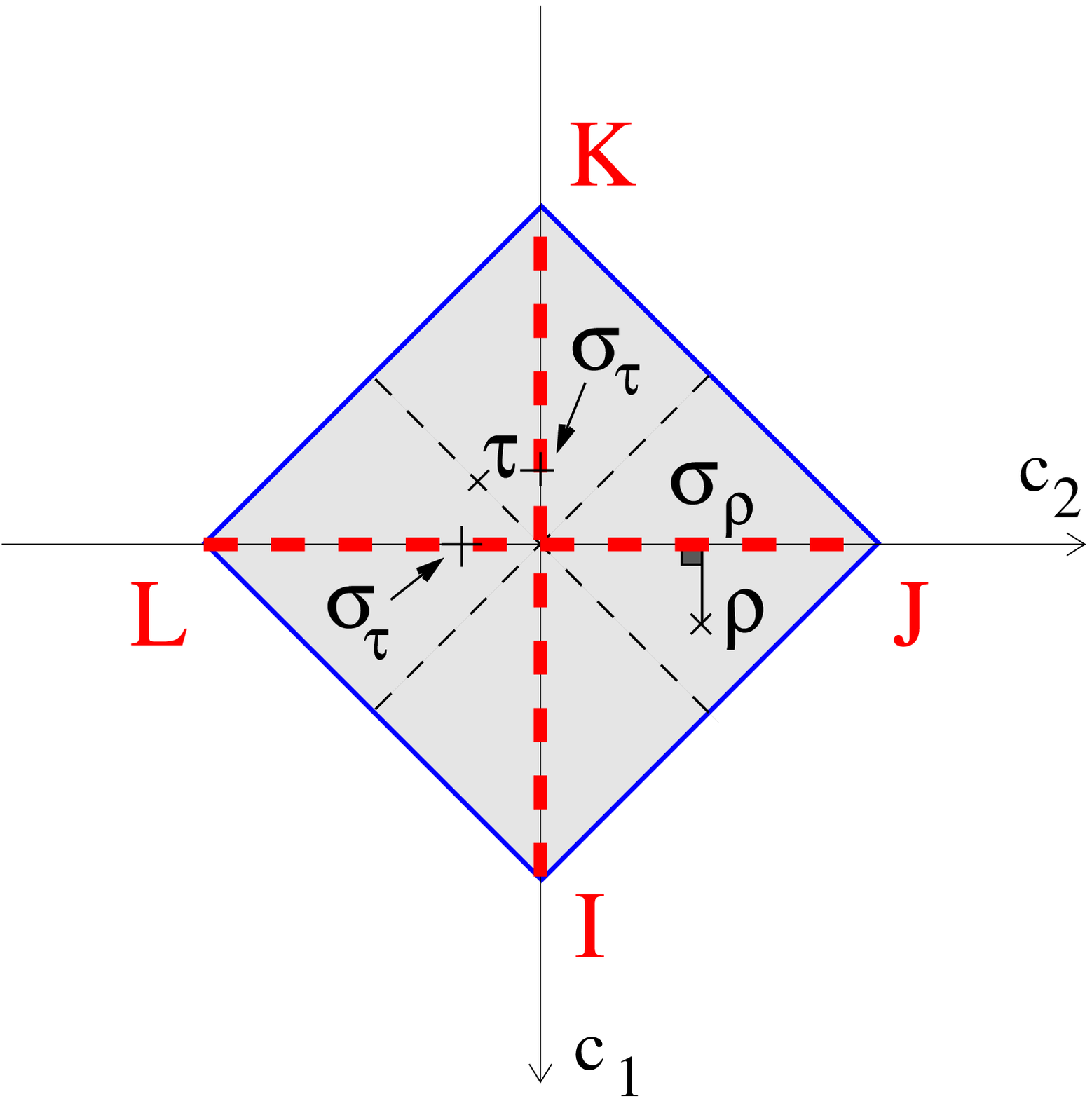}
\end{center}

\vspace*{-6mm}

\flushleft{\bf (c)} 

\vspace*{-7mm}

\begin{center}
\includegraphics[width=0.23\textwidth]{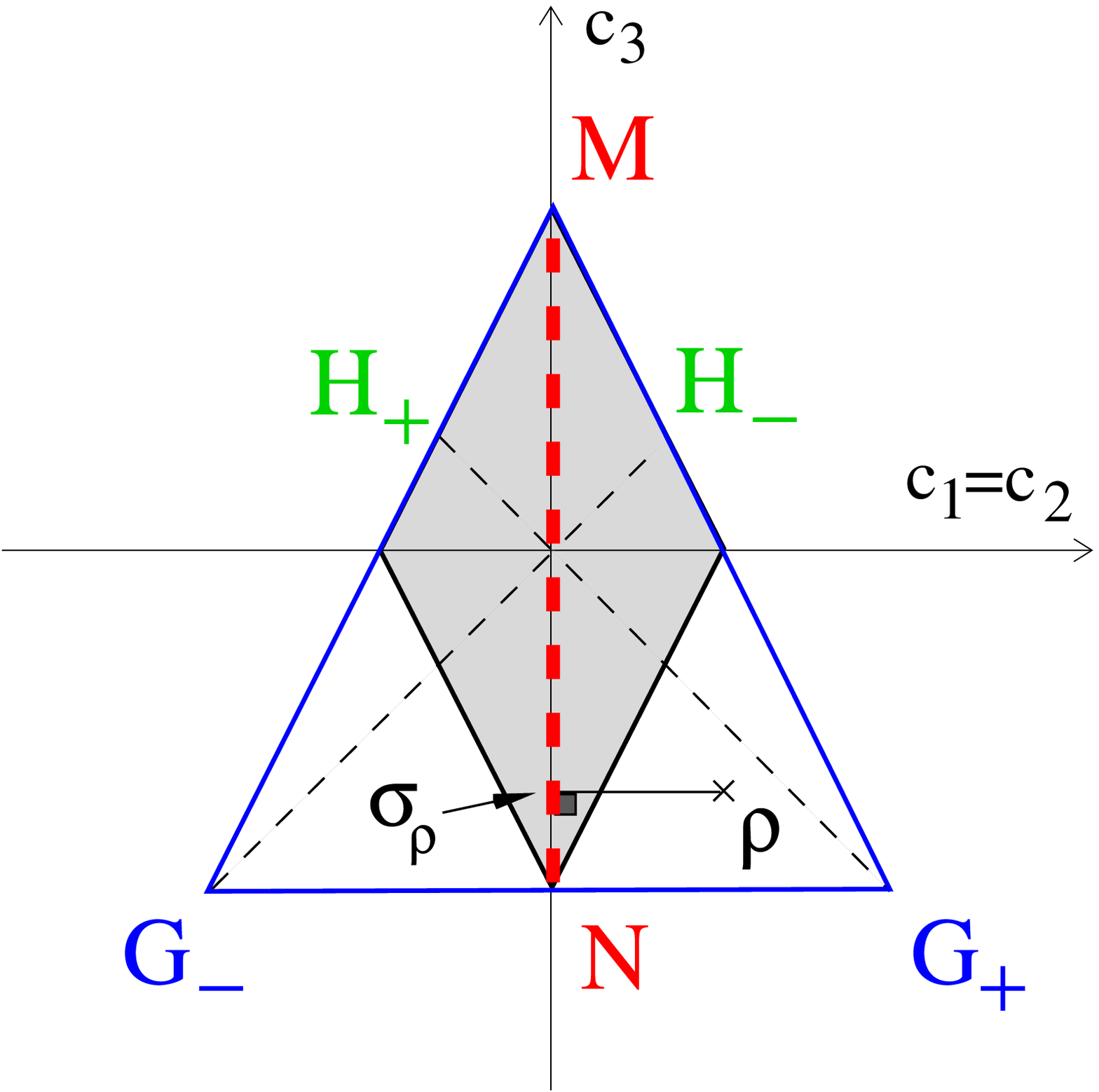}
\end{center}

\caption{(Color online) \label{fig3}
(a) Tetrahedron ${\cal T}$ with vertices $F_{\pm}=(\pm 1, \mp 1, 1)$ and $G_\pm = (\pm 1, \pm 1,-1)$,
and center $O(0,0,0)$.
The vectors $\cv \in {\cal T}$ represent physical states $\rho (\cv)$ with maximally 
mixed marginals.
The $A$-classical states in ${\cal T}$ are also $B$-classical and 
are located on the segments $[I K]$, $[J L]$, and $[M N]$  
(thick red dashed lines), with
$I(1,0,0)$, $J(0,1,0)$, $K(-1,0,0)$, $L(0,-1,0)$, $M(0,0,1)$, and $N(0,0,-1)$. 
The shaded region delimited by the octahedron $IJKLMN$ corresponds to separable states.
(b) Cut view of ${\cal T}$ in the plane $c_3=0$. The state $\rho(\frac{1}{4},\frac{1}{2},0)$
has a unique closest classical state (CCS) $\sigma_\rho (0,s_2,0)$ with 
$s_2 \simeq 0.523$
(see Eq.(\ref{eq-closest_state_interior_tetrahedron})).
States on the broken lines are at equal Bures distance from the segments $[IK]$ and $[JL]$ and
have infinitely many CCSs, given by Eq.(\ref{eq-additional_formula}).
On these lines $D_A(\rho)$ is not differentiable. Two CCSs 
of the form (\ref{eq-state_with_max_disord_marginals}) to
the state $\tau  (-\frac{1}{4},-\frac{1}{4},0)$,
corresponding to $\uv=\ev_1$ and $\ev_2$  in (\ref{eq-additional_formula}), are shown:  
$\sigma_\tau (s_1,0,0)$  and $\sigma_\tau ( 0,s_1,0)$ with
$s_1 \simeq - 0.259$.
(c) Cut view of ${\cal T}$ in the plane $c_1=c_2$. States in the shaded region are separable.
The state $\rho (\frac{1}{2},\frac{1}{2},-\frac{3}{4})$ has a unique CCS
$\sigma_\rho ( 0,0,s_3)$, with 
$s_3 \simeq -0.729$. 
States inside the triangles $H_{\mp } O G_\pm$ 
have infinitely many CCSs, given by (\ref{eq-additional_formula}).  
All states lying on the blue edges of the square and triangle in panels (b) and (c), and 
on the faces of ${\cal T}$ in panel (a),
have infinitely many CCSs, given by 
Eqs.(\ref{eq-closest_state_e_z}) and (\ref{eq-closest_state_e_zbis}).
}
\end{figure}

It is clear on (\ref{eq-b}) that
the vectors $\cv$  such that
$b_{\rm max}$ takes the highest possible value $b_{\rm max}=1$  have a single non-vanishing component $c_m$. 
Therefore, in agreement with the results of Ref.~\cite{Dakic10}, the $A$-classical 
states with maximally mixed marginals are located (up to local unitary equivalence) 
on the three segments of the coordinate axes inside the tetrahedron, represented 
in Fig.~\ref{fig3}(a) by thick dashed lines.
As a consequence, all  states $\rho$ such that $\cv$ is inside the octahedron  
formed by the convex hull of the three aforementioned segments
are separable (for indeed, convex combinations of $A$-classical states
are separable).
It follows from the Peres-Horodecki criterium that this octahedron contains all separable
states with maximally mixed marginals~\cite{Horodecki96}.

The states (\ref{eq-state_with_max_disord_marginals}) with the highest 
discord $\widetilde{D}_A=1$ have $b_1=b_2=b_3=0$.
These states are
the four Bell states $\kket{\Psi_\nu}$
located at the vertices 
$F_\pm$ and $G_\pm$ of the tetrahedron (see Fig.~\ref{fig3}(a)). Note that an analogous result holds for 
the Hilbert-Schmidt-GQD~\cite{Dakic10}.

Let us end this subsection by remarking that  for two-qubit states $\rho$ with maximally mixed marginals
$\rho_A=\rho_B=1/2$, the inequality (\ref{eq-inequality_square_root_meas}) is an equality. 
Actually, for such $\rho$ one has equal
 prior probabilities $\eta_i = \bbra{\alpha_i} \rho_A \kket{\alpha_i}=1/2$  in the
QSD task of Sec.~\ref{sec-link_D_A_QSD}. Moreover, it follows 
from (\ref{eq-state_with_max_disord_marginals}) 
that $\rho$ is invariant under conjugation by the spin-flip operators
$\sigma_1 \otimes \sigma_1$ and $\sigma_2 \otimes \sigma_2$. But it has been observed above that 
an optimal direction $\uv^\opt$ is given by one of the coordinate vectors $\ev_m$ or its opposite. 
Disregarding irrelevant phase factors, the two eigenvectors $\kket{\alpha_0^\opt}$
 and $\kket{\alpha_1^\opt}$ of $\sigma_{\uv^\opt}$ are  transformed one into another 
by the spin-flip operator $\sigma_m$, with $m=1$ if $\uv^\opt = \pm \ev_2, \pm \ev_3$ and/or 
$m=2$ if $\uv^\opt=\pm \ev_1, \pm \ev_3$. Hence the states $\rho_i$  are related by a unitary conjugation,
$\rho_1 = \sigma_m \otimes \sigma_m \,\rho_0 \, \sigma_m \otimes \sigma_m$.
The square-root measurement is known to be
optimal to discriminate such symmetric ensembles of states  with equal prior
probabilities~\cite{Eldar04,Chou03}. Thus $F_A (\rho)$ 
is given by the right-hand side  of Eq.(\ref{eq-inequality_square_root_meas}).
Based on this observation, one can rederive formula (\ref{eq-max_fidelity_and_GDQ_Luo}) 
in a slightly simpler way.
%
%

\subsection{Comparison with the other discords} \label{sec-comparison}

\begin{figure}[t]
\centering
\includegraphics[width=0.48\textwidth]{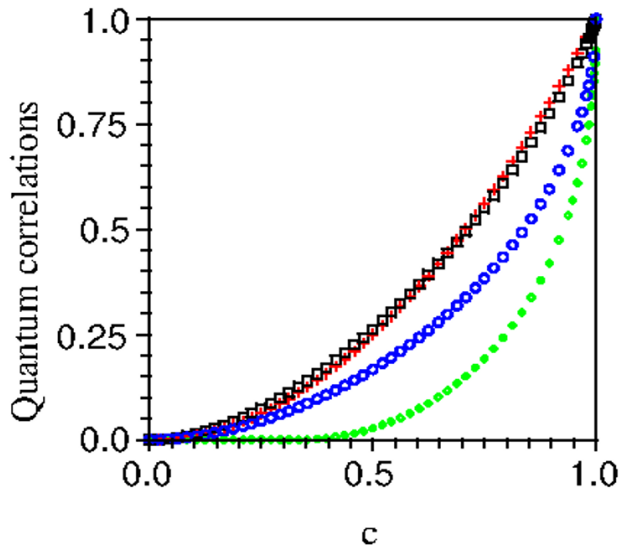}
\centering
\caption{\label{fig-fig4}
(Color online) Comparison of the normalized 
discords and geometric entanglement for a Werner state 
$\rho_W  = \frac{1-c}{4}\,  1 +c \ketbra{\Psi^{-}}{\Psi^{-}}$ with $c$ varying between $0$ and $1$ 
($\rho_W$ is 
located on the segment $[O G_-]$ in Fig.~\ref{fig3}(a)).
From top to bottom:
quantum discord $\delta_A$ (black boxes),
GQD $\widetilde{D}_A^{(2)}$ with  Hilbert-Schmidt distance (red crosses), 
GQD $\widetilde{D}_A$ with Bures distance (blue circles), and
geometric measure of entanglement $\widetilde{E}$ (green diamonds). 
}
\end{figure}

We can now  compare the normalized GQD $\widetilde{D}_A(\rho)$ for states $\rho$ of the form 
(\ref{eq-state_with_max_disord_marginals}) with other measures of quantum correlations.
The corresponding quantum discord
\begin{eqnarray} \label{eq-result_Luo}
\nn
\delta_A (\rho)&  =& \sum_{\nu=0}^3 p_\nu \ln_2 p_\nu + 2 - \frac{1-|\cv |}{2} \ln_2 ( 1-|\cv|) \\
& & - \frac{1+|\cv|}{2} \ln_2 ( 1 + |\cv|) 
\end{eqnarray}
has been calculated in~\cite{Luo08b}. Here,  $|\cv | = \max_m | c_m|$ and the probabilities $p_\nu$ are
given by (\ref{eq-expression_p_m}). 
The GQD with Hilbert-Schmidt distance, $D^{(2)}_A (\rho)= d_2 ( \rho, {\Cc}_A )^2 \equiv \min_{\sigma \in \Cc_A}
\tr [( \rho-\sigma )^2]$,  
is easy to determine for  arbitrary two-qubit states~\cite{Dakic10}.
For the states (\ref{eq-state_with_max_disord_marginals}) it reads
$D^{(2)}_A (\rho) = (\sum_{m} c_m^2 - |\cv |^2)/4$. Since the maximal value  of
$D^{(2)}_A$ is $1/2$, we normalize it as
$\widetilde{D}^{(2)}_A(\rho) = 2 D^{(2)}_A (\rho)$.
A third discord considered in~\cite{Modi10} is defined with the help of the relative entropy
$S(\rho||\sigma)= \tr ( \rho \ln \rho) - \tr ( \rho \ln \sigma)$ as
 $\Delta_A (\rho) = \min_{\sigma \in \Cc_A} S ( \rho || \sigma)$. 
However, for the states (\ref{eq-state_with_max_disord_marginals}) it
coincides with $\delta_A(\rho)$~\cite{Mazzola10}.
We also compare $\widetilde{D}_A(\rho)$ with the geometric measure of entanglement (see Sec.~\ref{sec-def_QD}),
given for two qubits by
$E(\rho)= 2 - \sqrt{2} (1 +  \sqrt{1-C(\rho)^2} )^{1/2}$ \cite{Streltsov10}
where $C(\rho)$ is the
 Wootters concurrence~\cite{Wootters98}. In our case 
$C(\rho)= \max\{ |c_1-c_2|-1+c_3, |c_1+c_2|- 1-c_3, 0\}/2$. 
The measure $E$ is normalized as $\widetilde{E}(\rho)=E(\rho)/(2-\sqrt{2})$.
Let us point out that $\widetilde{E}(\rho) \leq \widetilde{D}_A (\rho)$ for any $\rho$, since
$A$-classical states are separable.

In Fig.~\ref{fig-fig4}, the four quantum correlation measures $\widetilde{D}_A$, $\delta_A$, 
 $\widetilde{D}_A^{(2)}$, and $\widetilde{E}$ are plotted together
for the Werner states, obtained by taking $c_1=c_2=c_3=-c$ in (\ref{eq-state_with_max_disord_marginals}),
with $0 \leq c \leq 1$.

\begin{figure}
\begin{minipage}[t]{7cm}%
\centering
{\bf (a)}
\includegraphics[width=0.8\textwidth]{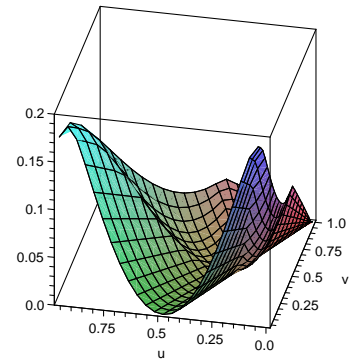}
\end{minipage}\hspace{1cm}
\begin{minipage}[t]{7cm}
\centering
{\bf (b)} 
\includegraphics[width=0.8\textwidth]{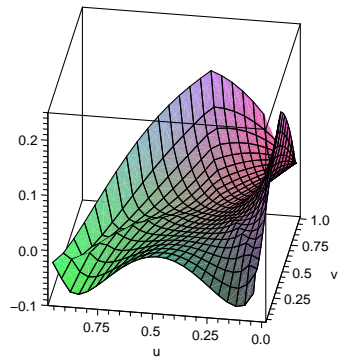}
\end{minipage}

\caption{(Color online) \label{fig-fig5}
(a) Difference $\delta_A(\rho)-\widetilde{D}_A(\rho)$ between the quantum discord
and the normalized Bures-GQD for  
states with 
maximally mixed marginals with vectors $\cv$ inside the triangle of vertices $F_+$,
$F_{-}$, and $N$ (see Fig.~\ref{fig3}(a)). 
The parameters $u$ and $v$, $0 \leq u \leq 1-v \leq 1$, 
are chosen such that $c_1=-2 u - v+1=-c_2 $ and $c_3=-2 v+1$; 
(b) same for the difference $\delta_A(\rho)- \widetilde{D}_A^{(2)}(\rho)$
between the quantum discord and the normalized Hilbert-Schmidt-GQD.}
\end{figure}

Fig.~\ref{fig-fig5}(a) displays the difference $\widetilde{D}_A(\rho)- \delta_A(\rho)$
for a subfamily of states with maximally mixed marginals depending on two parameters.
We observe that
\begin{equation} \label{eq-inequality_discords}
\delta_A(\rho) \geq \widetilde{D}_A (\rho)\;.
\end{equation}
In contrast, one sees in Figs.~\ref{fig-fig4} and~\ref{fig-fig5}(b) that the GQD with Hilbert-Schmidt distance
can be either smaller or larger than $\delta_A$, as it
has been noted previously in other works~\cite{Girolami11}.
We have looked numerically for vectors $\cv$ inside the tetrahedron
violating (\ref{eq-inequality_discords}) and have not found any such vector.
This indicates that this bound holds for any state $\rho$ with 
maximally mixed marginals.
It would be of interest to know if it also holds for more general
states. Let us recall from Sec.~\ref{sec-DQC1} that (\ref{eq-inequality_discords}) is true
for the output state of the DQC1 model with random unitaries.

\subsection{Closest $A$-classical states} 
\label{sec-closest_A_clas_state}

We now turn to the problem of finding the $A$-classical states 
$\sigma_\rho$  satisfying
\begin{equation} \label{eq-def_closest_state}
d_B(\rho,\sigma_\rho )^2 = D_A (\rho) = \min_{\sigma_\Aclass \in \Cc_A} d_B (\rho, \sigma_\Aclass )^2\;.
\end{equation}
These closest $A$-classical states to $\rho$
provide useful information on $\rho$ and on the geometry of the set  of quantum states.
Such information, which is not contained in the discord $D_A (\rho)$,  can be of interest when studying 
dynamical evolutions, for instance,  when the
two subsystems are coupled to their environments and undergo decoherence processes.
Like in the two previous subsections, we focus on two-qubit states with maximally mixed marginals 
given by Eq.(\ref{eq-state_with_max_disord_marginals}).
The geometric representation of these states
by vectors $\cv$ in the tetrahedron ${\mathcal T}$ will give some insight to our results.
In order to compare the geometries on ${\mathcal T}$ associated to the two distances $d_B$ and $d_2$, 
we also  determine the closest $A$-classical states for the Hilbert-Schmidt distance $d_2$
by using the results of Ref.~\cite{Dakic10}.

The states $\sigma_\rho$ are obtained by applying formula (\ref{eq-again_I_was_stupid}). 
In view of the results of Sec.~\ref{sec-2_qubit_general}, the optimal basis $\{ \kket{\alpha_i^\opt} \}_{i=0}^1$ 
is  the eigenbasis of the spin operator $\sigma_{\uv^\opt}$ in the optimal direction $\uv^\opt$. 
We  already know  that for the states (\ref{eq-state_with_max_disord_marginals}),
$\uv^\opt$ is parallel to the coordinate vector $\ev_{m_{\rm max}}$ if the maximum $| \cv | =
\max_{m=1,2,3} |  c_m|$ is reached for a single index $m =\mmax$ 
(see Sec.~\ref{seq-GDQ_for_states_with_max_mixed_marg}); 
otherwise, $\uv^\opt$ is an arbitrary
unit vector in the space spanned by the $\ev_m$ 
corresponding to indices $m$ such that $|c_m|=|\cv |$ (indeed, 
for such indices the parameters $b_m$ in (\ref{eq-max_fidelity_intermediate_step})
 are equal, $b_m=b_{\rm max}$, by virtue of (\ref{eq-order_c_m_and_b_m})).  
Moreover, it has been shown in Sec.~\ref{seq-GDQ_for_states_with_max_mixed_marg} 
that the states $\rho_i^\opt$ are optimally discriminated
by the square root measurement given by the projectors
\begin{equation} \label{eq-square_root_measurement_op} 
\Pi_i^\opt= \ketbra{\alpha_i^{\rm opt}}{\alpha_i^{\rm opt}} \otimes 1\;.
\end{equation}

Let us first take $\uv^\opt = \pm \ev_{\mmax}$. Then 
$\sigma_{\mmax} = \pm ( \ketbra{\alpha_0^\opt}{\alpha_0^\opt} - \ketbra{\alpha_1^\opt}{\alpha_1^\opt} )$.
One infers from the expression (\ref{eq-Luo_density_matrix2}) of the matrix
$\sqrt{\rho}$ in the standard basis that
$\sqrt{\rho} = \sum_{\nu=0}^3 t_\nu \sigma_\nu \otimes \sigma_\nu$, where 
$t_\nu$ are given by (\ref{eq-definition_t_nu}) and we have set $\sigma_0 = 1$.
This yields 
\begin{equation} \label{eq-intermediate_step_CCS_interior_T}
\bbra{\alpha_i^\opt} \sqrt{\rho} \kket{\alpha_i^\opt} = t_0 \pm (-1)^i t_{\rm max} \sigma_{\mmax}
\end{equation}
with $t_{\rm max} \equiv t_{\mmax}$. Replacing (\ref{eq-square_root_measurement_op}) and 
(\ref{eq-intermediate_step_CCS_interior_T})
into  (\ref{eq-again_I_was_stupid}) and  using the identities  
$F_A(\rho)= 4 (t_0^2+t_{\rm max}^2)$ and $16 t_0 t_m = a_m + c_m$, which follow respectively
from (\ref{eq-max_fidelity_and_GDQ_Luo}) and (\ref{eq-b_m_and_a_m})
and from (\ref{eq-definition_t_nu}), (\ref{eq-expression_p_m}), and (\ref{eq-b_m_as_function_of_c_m}),  
one finds
\begin{equation} \label{eq-closest_state_interior_tetrahedron}
\sigma_\rho = \frac{1}{4} \Bigl( 1 \otimes 1 +  \frac{a_{\mmax}+c_{\mmax}}{2 F_A(\rho)}\,
 \sigma_{\mmax} \otimes \sigma_{\mmax} \Bigr)
\end{equation}
where 
\begin{eqnarray}
& & a_1 
 =    2 (- \sqrt{p_0 p_1} +  \sqrt{p_2 p_3} ) = \frac{1}{2}\times 
\\ \nn
& & 
\Bigl( \sqrt{ (1 + c_1)^2  - (c_2 - c_3)^2}
- \sqrt{ (1 - c_1)^2 - (c_2 + c_3)^2} \Bigr)\;,
\end{eqnarray}
with analogous formulas for $a_2$ and $a_3$ obtained via a permutation of indices.
Hence, a two-qubit state $\rho$ with maximally mixed marginals always 
admits among its closest $A$-classical states 
a state with maximally mixed marginals, given by Eq.(\ref{eq-closest_state_interior_tetrahedron}).
This closest state has the following decomposition in terms of the four Bell states
\begin{equation} \label{eq-closest_state_interior_tetrahedronbis}
\sigma_\rho = \frac{1-q_{\mmax}}{2} \!\!\!\! \sum_{\nu =0,\mmax}\!\!\!\!\!\! \ketbra{\Psi_\nu}{\Psi_\nu}
 + \frac{q_{\mmax}}{2} \!\!\!\!\!\! \sum_{\nu \not=0,\mmax}\!\!\!\! \ketbra{\Psi_\nu}{\Psi_\nu} 
\end{equation} 
with
\begin{equation} \label{eq-def-q_m}
q_{m} = \frac{(t_0 + t_m)^2}{2(t_0^2+t_m^2)} = \frac{1}{2} + \frac{a_m + c_m}{2 b_m+2}\;.
\end{equation}
Interestingly, the closest states (\ref{eq-closest_state_interior_tetrahedronbis}) for the Bures distance
have  the same form as the closest states 
for the relative entropy.
Actually, the states $\sigma_\rho^\Delta$ satisfying $S ( \rho || \sigma_\rho^\Delta)
= \min_{\sigma \in \Cc_A} S (  \rho || \sigma)$ are given by 
Eq.(\ref{eq-closest_state_interior_tetrahedronbis})  but with different probabilities $\widetilde{q}_{\mmax}$ 
and $1-\widetilde{q}_{\mmax}$, given by the sums
of the two largest or the two smallest  probabilities $p_\nu$~\cite{Modi10}.
According to our notation, these two smallest (respectively largest) probabilities are $p_0$ and $p_{\mmax}$ 
if $|c_m|=|\cv|$ for a single index $m = \mmax$ and 
$p_0+p_{\mmax} < 1/2$ (respectively $p_0+p_{\mmax} > 1/2$).

If $|c_m|= |\cv |$ for two indices $m$, say, $|c_1|=|c_2|> |c_3|$ with $c_1=\pm c_2$, 
all the eigenbasis $\{ \kket{\alpha_{\phi,i}^\opt}  \}$ of the spin operators
$\sigma_\phi = \cos \phi\, \sigma_1 + \sin \phi \, \sigma_2$
maximize the success probability 
$P_S^{\,\rm{opt\,v.N.}} ( \{ \rho_i,\eta_i \})$ in (\ref{eq-variationnal_formula_bis}).
Hence $\rho$ has infinitely many closest $A$-classical states $\sigma_\rho$. These states 
can be determined by a simple generalization of 
the calculation leading to (\ref{eq-closest_state_interior_tetrahedron}), choosing 
for $\Pi_i^\opt$ the square-root measurement operators 
(\ref{eq-square_root_measurement_op}) associated to  
$\{ \kket{\alpha_{\phi,i}^\opt}  \}$.
Let us note that $c_1 = \pm c_2$ implies $t_1 = \pm t_2$ and recall the identities
$\kket{\alpha_{\phi,i}^\opt} = e^{- i \frac{\phi}{2} \sigma_3} \kket{\alpha_{0,i}^\opt}$ and
 $e^{\I  \frac{\phi}{2} \sigma_3} \sigma_\uv e^{-\I  \frac{\phi}{2} \sigma_3} = \sigma_{\uv'}$,
where $\uv'$ is related to $\uv$ by  a rotation around the $z$-axis with the angle $-\phi$.
We find  the following family of closest  $A$-classical states depending on the optimal vector
$\uv = u_1 \,\ev_1 + u_2 \,\ev_2$ with $u_1^2 + u_2^2 = 1$:
\begin{equation}
\label{eq-additional_formula}
\sigma_\rho ( \uv ) 
  =  \frac{1}{4}
\Bigl( 1 \otimes 1 
+ \frac{a_{\mmax}+c_{\mmax}}{2 F_A (\rho)} \, \sigma_\uv \otimes \sigma_{\uv(\cv)}
\Bigr)
\end{equation}

\noindent with $\mmax \in \{ 1, 2\}$ and $\uv (\cv) = c_{\mmax}^{-1} (c_1 u_1\,\ev_1 + c_2 u_2\, \ev_2)$
(note that $\sigma_\rho ( \uv )$ is independent of the choice of $\mmax$ in $\{1,2\}$ since
$a_2 + c_2 = \pm (a_1 + c_1)$). 
Similar results are obtained  by permutation of indices in the cases $|c_1|=|c_3|>|c_2|$ and $|c_2|=|c_3|>|c_1|$.
The $A$-classical states (\ref{eq-additional_formula}) have maximally mixed marginals,
although they are not of the form 
(\ref{eq-state_with_max_disord_marginals}) save for $u_1=0,\pm 1$ (in fact, $\sigma_\rho ( \uv )$
can be transformed into the state (\ref{eq-closest_state_interior_tetrahedron}) by an appropriate
 conjugation by a local unitary operator).
If $|c_1|=|c_2|=|c_3|$, any orthonormal basis  is  optimal
and one obtains a family of closest states  depending on 
the arbitrary optimal  unit vector $\uv \in \real^3$,  given by
(\ref{eq-additional_formula}) with
$\uv (\cv) = c_{\mmax}^{-1} ( c_1 u_1 \,\ev_1 + c_2 u_2 \,\ev_2 + c_3 u_3 \, \ev_3)$.

Let us now discuss a  more subtle point, which is of relevance if one wants to find {\it all}
$A$-classical states $\sigma_\rho$ satisfying (\ref{eq-def_closest_state}). 
Given an optimal basis $\{ \kket{\alpha_i^\opt}\}$, there is not guarantee that
the square-root measurement (\ref{eq-square_root_measurement_op}) is the only optimal measurement 
 maximizing $P_S (\{ \rho_i^\opt , \eta_i^\opt \})$.
Actually, it turns out that this is not the case if 
$\cv$ belongs to one of the faces of the tetrahedron (so that $p_0 p_1 p_2 p_3 =0$).
In this situation, we encounter in the appendix   
a whole family of optimal projectors $\Pi_i^\opt(r)$, depending
on a real parameter $r \in [-1,1]$. Let us emphasize that 
if two states of the form (\ref{eq-again_I_was_stupid}) with the same optimal basis $\{ | \alpha_i^{\rm opt} \rangle \}$ are closest
$A$-classical states to $\rho$ then, by convexity of the square Bures distance, any convex combination
of these states is also a closest $A$-classical state to $\rho$. Hence the set of optimal projectors must be
either infinite or reduced to one point.
To each measurement in this set one can associate a closest $A$-classical state by 
Eq.(\ref{eq-again_I_was_stupid}), which is determined in the appendix. Before 
giving the result, let us  introduce  
the product basis $\{ \kket{\alpha_i,\beta_j} = \kket{\alpha_i} \otimes \kket{\beta_j} \}_{i,j=0}^1$
of $\complex^2 \otimes \complex^2$  defined as follows:
(i) if $|c_m|$ is maximal for a single index $m =\mmax$, then 
$\kket{\alpha_i}=\kket{\beta_i}$ are
the eigenvectors of $\sigma_{\mmax}$;
(ii) if  $|c_m|$ is maximum for exactly two components $c_m$, then $\mmax$,
$\kket{\alpha_i}$, and $\kket{\beta_i}$ are defined by
\begin{widetext}
\begin{equation} \label{eq-alpha_and_beta}
\mmax
= 
\begin{cases}
1 
\\[0.5em]
3
\\[0.5em]
3
\end{cases}
,\;
\kket{\alpha_i}
 = 
\begin{cases}
\dss
e^{- \I \frac{\phi}{2} \sigma_3} \frac{\kket{0} + (-1)^i  \kket{1}}{\sqrt{2}}
\\[0.5em] \dss
e^{-\I \frac{\theta}{2} \sigma_2} \kket{i}
\\[0.5em] \dss
e^{\I \frac{\theta}{2} \sigma_1} \kket{i}
\end{cases}
 ,\;
\kket{\beta_i}
 = 
\begin{cases}
\dss 
e^{\mp \I \frac{\phi}{2} \sigma_3}  \frac{\kket{0}+ (-1)^i \kket{1}}{\sqrt{2}}
& \text{if $c_1=\pm c_2$, $|c_1|>|c_3|$}
\\[0.5em]
\dss
e^{\mp \I \frac{\theta}{2} \sigma_2} \kket{i}
&  \text{if $c_1=\pm c_3$, $|c_1| > |c_2|$}
\\[0.5em]
\dss
e^{\pm \I \frac{\theta}{2} \sigma_1} \kket{i}
&  \text{if $c_2=\pm c_3$, $|c_2| >|c_1|$}
\end{cases}
\end{equation}
for some arbitrary angles $\phi$ or $\theta  \in [0, 2 \pi[$; (iii) if 
$|c_1|=|c_2|=|c_3|$, that is, $c_1=\epsilon_2 c_2 = \epsilon_3 c_3$ with $\epsilon_{2,3} \in \{-1,1\}$,
then
\begin{equation} \label{eq-alpha_and_beta_bis}
\mmax =3 \;,\;
\kket{\alpha_i}  =   e^{-\I \frac{\phi}{2} \sigma_3} e^{-\I \frac{\theta}{2} \sigma_2} \kket{i}
\; ,\; 
\kket{\beta_i}  =  e^{-\I \epsilon_2 \frac{\phi}{2} \sigma_3} e^{-\I \epsilon_3 \frac{\theta}{2} \sigma_2} \kket{i}
\end{equation}
for arbitrary angles $\phi$ and $\theta \in [0,2\pi[$.
Then all the closest $A$-classical states to $\rho$ are of the following form:    
\begin{equation}  \label{eq-closest_state_e_z} 
\sigma_\rho (r) 
=  \frac{q_{\mmax}}{2}
\bigl[ \ketbra{\alpha_0 , \beta_0}{\alpha_0 ,\beta_0 }
 + \ketbra{\alpha_1, \beta_1 }{\alpha_1 ,\beta_1} \bigr]
+ \frac{1-q_{\mmax}}{2}
\bigl[  (1 + r) \ketbra{\alpha_0, \beta_1}{\alpha_0, \beta_1} 
+ (1-r) \ketbra{\alpha_1, \beta_0}{\alpha_1 , \beta_0}  \bigr]
\end{equation}
if $p_0 p_{m_{\rm max}}=0$ and $p_m >0\;\;\forall\; m\not= \mmax$, and
\begin{equation}  \label{eq-closest_state_e_zbis} 
\sigma_\rho (r) 
= \frac{q_{\mmax}}{2}
\bigl[ ( 1 + r) \ketbra{\alpha_0 , \beta_0}{\alpha_0 ,\beta_0 }
 + ( 1- r) \ketbra{\alpha_1, \beta_1 }{\alpha_1 ,\beta_1} \bigr]
+  \frac{1-q_{\mmax}}{2}
\bigl[  \ketbra{\alpha_0, \beta_1}{\alpha_0, \beta_1} 
+ \ketbra{\alpha_1, \beta_0}{\alpha_1 , \beta_0} \bigr]
\end{equation}
\end{widetext}
if $p_0 p_{m_{\rm max}}>0$ and $p_1 p_2 p_3 =0$. In these equations $r \in [-1,1]$ is arbitrary.
When $\cv$ is in the interior of the tetrahedron ${\cal T}$ (\ie, $p_0 p_1 p_2 p_3 >0$),
all the closest states are obtained by setting  $r=0$  in 
Eqs. (\ref{eq-closest_state_e_z}) and (\ref{eq-closest_state_e_zbis}).
One then recovers the previous results (\ref{eq-closest_state_interior_tetrahedron})
and (\ref{eq-additional_formula}). 
Therefore, $\rho$ has a unique closest $A$-classical state if and only if 
 $\cv$ is in the interior of ${\cal T}$  and the maximum $|\cv |$ is non-degenerate
(case (i)).
In the case (ii), $\rho$ admits a 
1-parameter  (or, if $\cv$ belongs to a face of ${\cal T}$, a 2-parameter)
family of closest $A$-classical states. 
In the case  (iii), this family is a 2-parameter (or, if $\cv$ belongs to a face of ${\cal T}$, 
a 3-parameter) family.

It should be noted that the states  (\ref{eq-closest_state_e_z}) and (\ref{eq-closest_state_e_zbis})  
do not have maximally mixed marginals save for $r=0$. This means that the states $\rho$ on the
faces of ${\cal T}$ have closest $A$-classical states outside the tetrahedron. 
Let us also emphasize that all the closest $A$-classical states $\sigma_\rho$ determined in this section 
are in fact classical states, that is, 
the eigenvectors of $\sigma_\rho$ are product states. Thus 
\begin{equation}
D_A (\rho) = D_B (\rho) = 2 \Bigl( 1 - \sqrt{\frac{1+ b_{\rm max}}{2}} \Bigr)\;,
\end{equation}
as could be expected from the symmetry of $\rho$ under
the exchange of the two qubits.
 
In Figs.~\ref{fig3}(b) and~\ref{fig3}(c), some examples of states $\rho$ in the tetrahedron
and their closest states $\sigma_\rho$  
are represented. In Fig.~\ref{fig3}(b), 
outside the dashed lines on which two or more components $c_m$ have equal moduli,
$\sigma_\rho$ is specified by a vector $\sv$ lying on the closest coordinate semi-axis to 
$\cv$ for the usual distance in $\real^3$ (but $\sv$ is  not the closest vector
to $\cv$ on that semi-axis).

The closest $A$-classical states for the Hilbert-Schmidt distance $d_2$
to a state $\rho$ of the form (\ref{eq-state_with_max_disord_marginals})
 can be found by using the results of Ref.~\cite{Dakic10}. They are given by
\begin{equation}
\sigma_\rho^{(2)}  (\uv)
 =  
  \frac{1}{4} \Bigl( 1\otimes 1 + \sum_{l,m=1}^3  u_l c_m  u_m \sigma_{l} \otimes \sigma_m \Bigr) 
\end{equation}
with an arbitrary unit vector $\uv \in \real^3$. In contrast with the Bures distance,
all states $\rho$ with maximally mixed marginals have infinitely many closest 
$A$-classical states. Moreover, there are three 
states $\sigma_\rho^{(2)}$ of the form (\ref{eq-state_with_max_disord_marginals})
 lying on the coordinate axes, with non-vanishing coordinate  equal to the corresponding
coordinate of  $\cv$,
irrespective of the order of the moduli $|c_m|$. 
Since $\tr [( \sigma_\rho^{(2)} )^2 ] = ( 1 + \sum_m c_m^2  u_m^2)/4 \leq 1/2$,
the closest states $\sigma_\rho^{(2)}$ are always mixed. In particular, unlike for the Bures distance, 
the closest classical states to the maximally entangled 
pure states $\kket{\Psi_\nu}$ are not pure product states.

\section{Conclusions} \label{sec-Conclusion}

We have shown that the results of Ref.~\cite{companion_paper} can be used  to determine 
the geometric quantum discord (GQD) with Bures distance 
of a state $\rho$ of a bipartite system $AB$ and
the corresponding closest $A$-classical states to $\rho$ when the subsystem $A$  is a qubit. 
Analytical expressions for this GQD
in the DQC1 model and for two qubits in a state with maximally mixed  marginals have been obtained.
In all cases for which exact analytical formulas for the usual quantum discord of Refs.~\cite{Ollivier01,Henderson01} 
are also available, we observe that the  normalized GQD is smaller than the usual discord. 
In the DQC1 model, the comparison of the GQDs of the output state 
for different unitaries $U_n$ shows that the highest discord appears when $U_n$
has  uniformly distributed eigenvalues on the unit disk modulo a symmetry
with respect to the origin. In particular, the discord cannot be larger than that
obtained for large random unitary matrices distributed according to the Haar measure, which
have been studied in~\cite{Datta08}. For rotation matrices and large number of qubits $n$,
the GQD is close to this upper bound, expected for small rotation  angles and at certain  
specific angles.  
For two-qubit states $\rho$  with maximally mixed marginals, 
the closest $A$-classical states for the Bures distance are qualitatively similar to the closest
$A$-classical states for the relative entropy, but completely different from those for the Hilbert-Schmidt 
distance.
The geometry  induced by the Bures distance in the tetrahedron looks like the Euclidean geometry of $\real^3$, 
the $A$-classical states being located on the three
segments of the coordinate axes inside the tetrahedron. However, 
depending on the symmetry of  $\rho$, $\rho$   
 can have either a unique closest classical state with maximally mixed marginals or a
continuous family of closest classical states with maximally mixed or non-maximally mixed marginals. 

Our results constitute a first step in the study  of the geometric measure
of quantum discord introduced in~\cite{companion_paper} in some concrete models
and its relation with the usual discord and other geometrical versions. 
It would be of interest to calculate the Bures-GQD 
in other physical models, in particular in the presence of dynamical evolutions, 
and to compare it with other measures of non-classicality in the literature.

\vspace{5mm}


\noindent {\bf Acknowledgements:}
We acknowledge financial support from
the French project no.\,ANR-09-BLAN-0098-01, 
the Chilean Fondecyt project no.\,100039, and the project Conicyt-PIA anillo no.\,ACT-1112 
``Red de An\'alisis estoc\'astico y aplicaciones''.

\vspace{4mm}

\noindent {\bf Note added:} After the completion of this work we have been informed 
by G. Adesso that the Bures geometric discord
for two-qubit states with maximally mixed marginals has been calculated independently
in~\cite{Aaronson13} by a different method, yielding to  
the same result (Eq. (\ref{eq-max_fidelity_and_GDQ_Luo})).

\appendix
\renewcommand{\theequation}{\Alph{section}\arabic{equation}}
\setcounter{equation}{0}
\section{Closest classical states of a two-qubit state with maximally mixed marginals} \label{app-A}

In this appendix we determine all optimal projective measurements  $\{ \Pi_i^\opt\}$   
 maximizing the success probability $P_S (\{ \rho_i^\opt , \eta_i^\opt \})$
in Eq.(\ref{eq-opt_success_proba}) for a two-qubit state $\rho$ 
of the form (\ref{eq-state_with_max_disord_marginals}). This allows us to find all the closest $A$-classical states
to $\rho$ by applying formula (\ref{eq-again_I_was_stupid}).

Before starting the calculation, let us first note that we can restrict our analysis to optimal 
directions $\uv^\opt$ satisfying $\uv^\opt \cdot \ev_m \geq 0$
for a fixed coordinate index $m=1,2$, or $3$.
Indeed, changing the order of the vectors in the basis $\{ \kket{\alpha_i^\opt} \}_{i=0}^1$ 
in Eqs.(\ref{eq-state_Q_discrimination}) and (\ref{eq-again_I_was_stupid})
amounts to exchange $(\rho_0^\opt,\eta_0^\opt) \leftrightarrow (\rho_1^\opt,\eta_1^\opt)$ and 
$\Pi_0^\opt \leftrightarrow \Pi_1^\opt$. 
This does clearly not modify the optimal success probability and the $A$-classical 
states $\sigma_\rho$. But $\{ \kket{\alpha_i^\opt} \}$ is the
eigenbasis of the spin operator $\sigma_{\uv}$ for the optimizing direction $\uv=\uv^\opt$, 
see Sec.~\ref{sec-2_qubit_general}.
Changing the order of the basis vectors thus corresponds to invert $\uv^\opt$. Hence we may assume
 $\uv^\opt \cdot \ev_m \geq 0$ without loss of generality.
Thanks to the results of Sec.~\ref{sec-GDQ_Luo_states}, we know that 
if $|c_m|$ is maximal for a single index $m = \mmax$ 
then $\uv^\opt= \ev_{m_{\rm max}}$. 

We first study the case $|c_3|>|c_1|,|c_2|$.
All other cases will be deduced by symmetry arguments.
Then $\uv^{\rm opt} = \ev_3$ and $\kket{\alpha_i^{\rm opt}} = \kket{i}$ ($i=0,1$) are the vectors of  
the standard basis.  
By  setting $\theta=0$ in (\ref{eq-Lambda_Luo_state}), 
we immediately find that $\Lambda^\opt \equiv \Lambda( \uv^\opt)$ has eigenvalues
$\lambda_+^\opt =(b_3+|a_3|)/4 >0$, 
$\lambda_-^\opt =(b_3-|a_3|)/4 \geq 0$, $-\lambda_{-}^\opt \leq 0$, and $-\lambda_{+}^\opt < 0$
(here, we have used the inequalities $b_3 \geq |a_3|$ and $b_3>0$, 
which follow from (\ref{eq-b_m_as_function_of_c_m})).
As stressed in Sec.~\ref{sec-2_qubit_general}, $\Pi_0^\opt$ 
is the projector on the direct sum $\Vv_+ \oplus \Vv_-$  of the eigenspaces $\Vv_{+}$ and $\Vv_{-}$ 
 associated to the maximal eigenvalues $\lambda_+^\opt$ and $\lambda_-^\opt$, provided that these eigenvalues are 
non-degenerated. If $\lambda_{-}^\opt =0$ is two-fold degenerated, 
any projector on $\Vv_+ \oplus \Vv$ with $\Vv \subset \Vv_{-}$  a one-dimensional 
subspace  of $\Vv_{-}$ defines an optimal projector $\Pi_0^\opt$.
In view of the diagonal form (\ref{eq-Lambda_Luo_state}) of $\Lambda^\opt$ in the standard basis, one gets 
\begin{equation} \label{eq-optimal_projector_qubits}
\Pi_0^\opt =
\begin{cases}
\; \ketbra{0}{0} \otimes 1 & \text{if $b_3 > | a_3|$}
\\
\; \ketbra{00}{00} + \ketbra{\Phi}{\Phi} &   \text{if $b_3=a_3>0$}
\\
\; \ketbra{01}{01} + \ketbra{\Phi'}{\Phi'} &  \text{if $b_3=-a_3>0$}      
\end{cases}
\end{equation}
with $\kket{\Phi} \in \Span \{ \kket{10}, \kket{01}\} $ and $\kket{\Phi'} \in \Span \{ \kket{00}, \kket{11} \}$,
$\| \Phi \| = \| \Phi'\|=1$.
The condition $b_3>|a_3|$ is achieved when all $p_\nu$ are nonzero, see (\ref{eq-b_m_as_function_of_c_m}).
Then the optimal measurement  $\{ \Pi_i^\opt\}$ is unique and thus $\rho$ has a unique  closest $A$-classical
state. The condition $b_3 = a_3$ (respectively $b_3 = -a_3$)
corresponds to $p_0 p_3=0$ (respectively $p_1 p_2=0$),
that is, to a vector $\cv$ belonging to the faces $F_+ F_- G_+$ or  
$F_+ F_- G_-$  (respectively, $G_+ G_- F_+$ or $G_+ G_- F_-$) of the tetrahedron.
In this situation one has infinitely many optimal measurements.
Note that $p_0 p_3$ and $p_1 p_2$ cannot
both vanish, since otherwise one would have $b_3=a_3 = 0$, in contradiction with
our hypothesis $|c_3|>|c_1|,|c_2|$ (which is equivalent to $b_3 > b_1,b_2\geq 0$
by (\ref{eq-order_c_m_and_b_m})).
Let us replace $\kket{\alpha_i^\opt}=\kket{i}$ and (\ref{eq-optimal_projector_qubits}) into the expression
(\ref{eq-again_I_was_stupid}) of $\sigma_\rho$ and make use of
the expression (\ref{eq-Luo_density_matrix2}) of $\sqrt{\rho}$ in the standard basis. 
By taking advantage of the identities 
$t_1+t_2=\pm (t_0-t_3)$ for $p_{0,3} =0$ and $t_1-t_2=\pm (t_0+t_3)$ for $p_{1,2} =0$
(which can be established with the help of (\ref{eq-definition_t_nu})), a
simple lengthly calculation yields
\begin{eqnarray} \label{eq-case_p_0p_3>0}
\nn
\sigma_\rho (r)
& = & 
\frac{(t_0+t_3)^2}{F_A(\rho)} \bigl[ \ketbra{00}{00} + \ketbra{11}{11} \bigr]
+ \frac{(t_0-t_3)^2}{F_A(\rho)} 
\\
& & \times \bigl[ (1+r) \ketbra{01}{01} + (1-r) \ketbra{10}{10} \bigr]
\end{eqnarray}
 if  $p_0 p_3 = 0$, and 
\begin{eqnarray} \label{eq-case_p_1p_2>0}
\nn
\sigma_\rho (r)
& = & 
\frac{(t_0+t_3)^2}{F_A(\rho)} \bigl[ (1+r) \ketbra{00}{00} + (1-r) \ketbra{11}{11} \bigr]
\\
& & + \frac{(t_0-t_3)^2}{F_A(\rho)}  \bigl[ \ketbra{01}{01} + \ketbra{10}{10} \bigr]
\end{eqnarray}
 if  $p_1 p_2 = 0$. The real parameter $r$ in (\ref{eq-case_p_0p_3>0}) and (\ref{eq-case_p_1p_2>0})
are given by 
$r= \pm 2  \re \{ \braket{01}{\Phi} \braket{\Phi}{10} \}$ 
and  
$r= \pm 2 \re \{ \braket{11}{\Phi'} \braket{\Phi'}{00} \}$, respectively.
Since $\kket{\Phi}$ and $\kket{\Phi'}$ are arbitrary normalized vectors in the two-dimensional spaces 
specified after (\ref{eq-optimal_projector_qubits}), $r$ can take any values in $[-1,1]$. 
If all $p_\nu$ are nonzero, the unique closest classical state is given by
setting $r=0$ in (\ref{eq-case_p_0p_3>0}) or (\ref{eq-case_p_1p_2>0}).
Recalling that $F_A (\rho) = (1+b_3)/2 =4 (t_0^2 + t_3^2)$, we find that
$\sigma_\rho$ is given by Eqs.(\ref{eq-closest_state_e_z}), (\ref{eq-closest_state_e_zbis}), and
(\ref{eq-def-q_m})
with $\mmax =3$ and $\kket{\alpha_i} = \kket{\beta_i} = \kket{i}$.

The case $|c_1|> |c_2|,|c_3|$ (respectively $|c_2| > |c_1|, |c_3|$) can be deduced from the
previous case by the  following symmetry argument. By means of the unitary conjugation 
$\rho'=U \rho \,U^\dagger$ with
$U=e^{\I \frac{\pi}{4} \sigma_2} \otimes  e^{\I \frac{\pi}{4} \sigma_2}$
(respectively $U=e^{-\I \frac{\pi}{4} \sigma_1} \otimes  e^{-\I \frac{\pi}{4} \sigma_1}$), one can transform 
$\rho$ into a state $\rho'$ 
of the form (\ref{eq-state_with_max_disord_marginals}) with a vector $\cv '=(c_3,c_2,c_1)$ 
(respectively $\cv '=(c_1,c_3,c_2)$) satisfying $|c_3'| > |c_1'|, |c_2'|$. 
By invariance of the fidelity (\ref{eq-fidelity}) and of $\Cc_A$ under local unitary conjugations,
such a transformation does not change $F_A$, so that 
$F_A (\rho)=F_A(\rho') = (1+b_3')/2=(1+b_{\rm max})/2$, 
in agreement with (\ref{eq-max_fidelity_and_GDQ_Luo}). 
Moreover, according to (\ref{eq-def_closest_state}), the states $\sigma_\rho$ 
are related to  the closest states $\sigma_{\rho'}$ to $\rho'$ by 
$\sigma_\rho = U^\dagger \sigma_{\rho'} U$. But $\sigma_{\rho'}$ is 
given by (\ref{eq-case_p_0p_3>0}) and (\ref{eq-case_p_1p_2>0})
 upon the replacement of $t_3$ by $t_3'=t_{{\rm max}}$. Therefore,
 $\sigma_\rho$ is given by Eqs.(\ref{eq-closest_state_e_z}) and (\ref{eq-closest_state_e_zbis}) 
in which $\kket{\alpha_i} = \kket{\beta_i}$ are the eigenvectors of $\sigma_{\mmax}$ with 
eigenvalues $(-1)^i$, that is,
$\kket{\alpha_i}= e^{-\I \frac{\pi}{4} \sigma_2} \kket{i} \propto \kket{0}  + (-1)^i \kket{1}$ if  
$|c_1|>|c_2|,|c_3|$ and
 $\kket{\alpha_i}= e^{\I \frac{\pi}{4} \sigma_1} \kket{i} \propto \kket{0}  + \I (-1)^i \kket{1}$ 
if  $|c_2|>|c_1|,|c_3|$. 
 
We now turn to states $\rho$ with vectors $\cv$ such that $|c_m|$ is maximum
for exactly two components $c_m$.
 For instance, let us assume $c_1=\pm c_2$ and $|c_1|=|c_2|> |c_3|$ 
(the other cases will easily follow  by a state rotation  
as above). Then, as noted in Sec.~\ref{sec-closest_A_clas_state},
 any vector $\uv_\phi= \cos  \phi \,\ev_1 + \sin \phi\, \ev_2$ 
in the $(xOy)$-plane defines an optimal direction. The corresponding optimal basis is
formed by the eigenvectors $\kket{\alpha_{\phi,i}^\opt} = e^{-\I \frac{\phi}{2} \sigma_3} \kket{\alpha_{0,i}^\opt}$
of $\sigma_{\uv_\phi}$ with eigenvalues $(-1)^i$.
The non-uniqueness of $\uv^\opt$ comes from the symmetry  of $\rho$. Actually, one finds
from (\ref{eq-state_with_max_disord_marginals}) that for any angle $\phi\in [0,2\pi[$,
\begin{equation}
\rho = U_\pm (\phi) \rho\, U_\pm (-\phi)
\; , \; U_\pm (\phi)= e^{-\I \frac{\phi}{2} \sigma_3} \otimes  e^{\mp \I \frac{\phi}{2} \sigma_3}\;.
\end{equation}
By using (\ref{eq-Lambda})  and the identity
$e^{\I  \frac{\phi}{2} \sigma_3} \sigma_\uv e^{-\I  \frac{\phi}{2} \sigma_3} = \sigma_{\uv'}$,
where $\uv'$ is related to $\uv$ by  a rotation around the $z$-axis with the angle $-\phi$,
 it follows that
\begin{equation}
\Lambda (  \uv_\phi ) = U_\pm (\phi) \Lambda ( \ev_1 ) U_\pm (-\phi)
\;.
\end{equation}
As a result, the spectral projectors $\Pi_{\phi,i}^\opt$ of $\Lambda (  \uv_\phi)$ 
and thus the
closest classical state (\ref{eq-again_I_was_stupid}) corresponding to $\uv^\opt=\uv_\phi$
are obtained from the corresponding projectors and  
classical state for $\uv^\opt=\ev_1$ by a unitary conjugation by $U_\pm (\phi)$.
But the closest classical state for $\uv^\opt=\ev_1$ is given by (\ref{eq-case_p_0p_3>0}) and 
(\ref{eq-case_p_1p_2>0}) upon the substitution of $t_3$ by $t_1$ and of $\kket{0},\kket{1}$ by  the eigenvectors of $\sigma_1$.   
We conclude that the states $\sigma_\rho$  are given
 by (\ref{eq-closest_state_e_z}) and (\ref{eq-closest_state_e_zbis})
with $\mmax=1$ and the vectors $\kket{\alpha_i}$ and $\kket{\beta_i}$ as in Eq.(\ref{eq-alpha_and_beta}).

Finally, let us study the states $\rho$ with  $|c_1|=|c_2|=|c_3|$.
This is the case for instance for Werner states. 
Let $\epsilon_{2} , \epsilon_{3} \in \{ -1,1\}$ be defined by $c_1 = \epsilon_2 c_2 =\epsilon_3 c_3$.
Then $\rho$ is invariant under rotations with arbitrary angles $\theta$ and $\phi$,
\begin{equation}
\rho = U_{\epsilon_2,\epsilon_3}  (\theta,\phi) \rho\, U_{\epsilon_2,\epsilon_3}  (\theta,\phi)^\dagger
\end{equation}
with
\begin{equation}
U_{\epsilon_2,\epsilon_3} (\theta,\phi) = e^{-\I \frac{\phi}{2} \sigma_3} e^{-\I \frac{\theta}{2} \sigma_2}
\otimes   e^{-\I \epsilon_2 \frac{\phi}{2} \sigma_3} e^{-\I \epsilon_3 \frac{\theta}{2} \sigma_2}\;.
\end{equation}
Moreover, the eigenvalues  of $\Lambda (\uv)$ are independent of $\uv$,
since $b_1=b_2=b_3$, see (\ref{eq-eigenvalues_Lambda}) and (\ref{eq-order_c_m_and_b_m}). 
Thus the optimal direction $\uv^\opt$ is
completely arbitrary and any orthonormal basis $\{ \kket{\alpha_i^\opt} \}$
of $\complex^2$ is optimal. The closest classical states to $\rho$ are 
again given by (\ref{eq-closest_state_e_z}) and (\ref{eq-closest_state_e_zbis}), with
the vectors $\kket{\alpha_i}$ and $\kket{\beta_i}$ as in Eq.(\ref{eq-alpha_and_beta_bis}).



\begin{thebibliography}{33}
\bibitem{remark1} In the literature these states are often referred to as the ``classical-quantum'' states
\bibitem{remark0} More precisely, a pure state can have either a unique closest separable state given 
by a pure product state or, if the maximal Schmidt coefficient $\mu_{\rm max}$ is degenerate, infinitely many 
closest separable states, given by convex combinations of orthogonal pure  product states
(see~\cite{companion_paper}).
\bibitem{remark2} Moreover, the ``most quantum''  states $\rho$
having a minimal fidelity $F_A(\rho)$ are the maximally entangled pure states
or convex combinations of such states with reduced $B$-states having supports on orthogonal subspaces.
\end{thebibliography}
\end{document}